\documentclass[sigconf]{acmart}
\usepackage{lscape}
\usepackage{dirtytalk}
\usepackage{booktabs}

\AtBeginDocument{%
  \providecommand\BibTeX{{%
    \normalfont B\kern-0.5em{\scshape i\kern-0.25em b}\kern-0.8em\TeX}}}

\copyrightyear{2023}
\acmYear{2023}
\setcopyright{acmlicensed}\acmConference[ASSETS '23]{The 25th International ACM SIGACCESS Conference on Computers and Accessibility}{October 22--25, 2023}{New York, NY, USA}
\acmBooktitle{The 25th International ACM SIGACCESS Conference on Computers and Accessibility (ASSETS '23), October 22--25, 2023, New York, NY, USA}
\acmPrice{15.00}
\acmDOI{10.1145/3597638.3608409}
\acmISBN{979-8-4007-0220-4/23/10}
\acmConference[ASSETS '23]{The 25th International ACM SIGACCESS Conference on Computers and Accessibility}{October 22--25, 2023}{New York, NY, USA}
\acmBooktitle{The 25th International ACM SIGACCESS Conference on Computers and Accessibility (ASSETS '23), October 22--25, 2023, New York, NY, USA}
\acmPrice{15.00}
\acmISBN{979-8-4007-0220-4/23/10}
\begin{document}

%%
%% The "title" command has an optional parameter,
%% allowing the author to define a "short title" to be used in page headers.
\title[Understanding Body Movement Education of BLV]{Understanding Challenges and Opportunities in Body Movement Education of People who are Blind or have Low Vision}

%%
%% The "author" command and its associated commands are used to define
%% the authors and their affiliations.
%% Of note is the shared affiliation of the first two authors, and the
%% "authornote" and "authornotemark" commands
%% used to denote shared contribution to the research.
%\author{Author 1}
%\authornote{}
%\email{}
%\orcid{1234-5678-9012}
%\affiliation{%
%  \institution{}
%  \streetaddress{}
%  \city{}
%  \state{}
%  \country{}
%  \postcode{}
%}

\author{Madhuka De Silva}
\orcid{0000-0002-3110-0990}
\affiliation{%
  \institution{Monash University}
  %\streetaddress{P.O. Box XXXX}
  %\city{Clayton}
  %\state{Victoria}
  \country{Australia}
  %\postcode{XXXX}
}
\email{madhuka.desilva@monash.edu}

\author{Sarah Goodwin}
\orcid{0000-0001-8894-8282}
\affiliation{%
  \institution{Monash University}
  %\streetaddress{P.O. Box XXXX}
  %\city{Clayton}
  %\state{Victoria}
  \country{Australia}
  %\postcode{XXXX}
}
\email{sarah.goodwin@monash.edu}

\author{Leona Holloway}
\orcid{0000-0001-9200-5164}
\affiliation{%
  \institution{Monash University}
  %\streetaddress{P.O. Box XXXX}
  %\city{Clayton}
  %\state{Victoria}
  \country{Australia}
  %\postcode{XXXX}
}
\email{leona.holloway@monash.edu}

\author{Matthew Butler}
\orcid{0000-0002-7950-5495}
\affiliation{%
  \institution{Monash University}
  %\streetaddress{P.O. Box XXXX}
  %\city{Clayton}
  %\state{Victoria}
  \country{Australia}
  %\postcode{XXXX}
}
\email{matthew.butler@monash.edu}

%\author{Author 2}
%\affiliation{%
%  \institution{}
%  \city{}
%  \country{}}
%\email{}

%\author{Author 3}
%\affiliation{%
%  \institution{}
%  \city{}
%  \country{}}
%\email{}

%\author{Author 4}
%\affiliation{%
%  \institution{}
%  \city{}
%  \country{}}
%\email{}

%%
%% By default, the full list of authors will be used in the page
%% headers. Often, this list is too long, and will overlap
%% other information printed in the page headers. This command allows
%% the author to define a more concise list
%% of authors' names for this purpose.
%\renewcommand{\shortauthors}{Author 1, et al.}

%%
%% The abstract is a short summary of the work to be presented in the
%% article.
%TC:ignore
\begin{abstract}
Actively participating in body movement such as dance, sports, and fitness activities is challenging for people who are blind or have low vision (BLV). Teachers primarily rely on verbal instructions and physical demonstrations with limited accessibility. Recent work shows that technology can support body movement education for BLV people. However, there is limited involvement with the BLV community and their teachers to understand their needs. By conducting a series of two surveys, 23 interviews and four focus groups, we gather the voices and perspectives of BLV people and their teachers. This provides a rich understanding of the challenges of body movement education. We identify ten major themes, four key design challenges, and propose potential solutions. We encourage the assistive technologies community to co-design potential solutions to these identified design challenges promoting the quality of life of BLV people and supporting the teachers in the provision of inclusive education.
\end{abstract}
%TC:endignore

%%
%% The code below is generated by the tool at http://dl.acm.org/ccs.cfm.
%% Please copy and paste the code instead of the example below.
%%


%%
%% Keywords. The author(s) should pick words that accurately describe
%% the work being presented. Separate the keywords with commas.

%% A "teaser" image appears between the author and affiliation
%% information and the body of the document, and typically spans the
%% page.
% \begin{teaserfigure}
%   \includegraphics[width=\textwidth]{figures/sampleteaser.pdf}
%   \caption{Seattle Mariners at Spring Training, 2010.}
%   \Description{Enjoying the baseball game from the third-base
%   seats. Ichiro Suzuki preparing to bat.}
%   \label{fig:teaser}
% \end{teaserfigure}

%%
%% This command processes the author and affiliation and title
%% information and builds the first part of the formatted document.
\maketitle

\section{Introduction}
 Accessing body movement educational programs in recreational and physical activities, such as sports, exercise, martial arts, and dance, is challenging for people who are blind or have low vision (BLV). This impacts the fitness of BLV people ~\cite{haegele2019, capella2007}, contributing to higher levels of sedentary behaviours ~\cite{starkoff2016}, becoming less actively involved with the world ~\cite{seham2015,lieberman2002}, and having less aesthetic and physical literacy. This can result in lower quality of life than people who are sighted~\cite{seham2015, timmons2019, holbrook2009}. Teachers of body movement primarily educate their students by instructing them to copy their demonstrations ~\cite{duggar1968}. However, visual cues play little or no role in educating BLV people in body movement. Instead, they mostly rely on verbal instructions provided by the teachers ~\cite{seham2012}. Verbal instructions tend to be ambiguous ~\cite{seham2012, esatbeyoglu2021}, and the style of instruction creates friction in understanding the movement ~\cite{bisset2016}. Physical interactions or touch demonstrations are also used in addition to verbal instructions when educating BLV people in body movement~\cite{duquette2012, oConnell2006, seham2012}. However, physical guidance and touch can lead to misinterpretation and are unsuitable for students sensitive to touch ~\cite{oConnell2006, holloway2022}. In addition to formal face-to-face educational methods, there has been a recent rise in videos and other online content for physical activity programs. This has been particularly evident in recent years, where we saw a huge increase in remote learning globally due to the Covid-19\footnote{COVID-19 is a highly infectious respiratory illness caused by the SARS-CoV-2 virus that first emerged in December 2019 and has since caused a global pandemic with substantial impacts on public health, society, and the economy.} lockdowns ~\cite{brinsley2021, brosnan2021, mcDonough2022}. However, it is questionable as to what extent remote content supports BLV people's learning, as most of the content is of a visual nature. 

Several HCI and accessibility research studies have been conducted aiming to support BLV people’s participation in body movement activities such as sports ~\cite{miura2018, watanabe2022, cooper2022, aggravi2016, muehlbradt2017}, and exercise ~\cite{malik2021, morellivi-bowling2010, morellivi-tennis2010, rector2017}, but fewer in creative body movement activities such as dance ~\cite{dias2019}. Some of those studies support the learning aspect by contributing to understanding body movement concepts and ensuring making proper adjustments~\cite{rector2017, dias2019}, whereas a majority address assistance and training needs that enhance the performance of the already learned concepts or techniques~\cite{miura2018}. Further, only a few studies have consulted teachers of particular activities ~\cite{rector2017, dias2019, cooper2022, bandukda2021}, with little to no focus on addressing challenges experienced by teachers when educating BLV people. Another gap is that only some solutions are based on an initial exploration of the needs of BLV people ~\cite{rector2015, india2021}, and little is known about how BLV people or their teachers adopt technology driven tools in body movement education. 

To address these gaps and understand the problem space following the recommendation to ``start with a problem instead of the technology'' ~\cite{siu2019}, we conducted a series of three studies with BLV adults and teachers who have experience in educating BLV people: a preliminary survey; semi-structured interviews; and focus group discussions. We aimed to answer two main research questions: (1) What are the current body movement learning methods, teaching techniques and tools for educating BLV people? and (2) What are the challenges and priorities of access to learning or teaching body movement for BLV people?
%The preliminary surveys \textbf{(Section 4)} were briefly capturing the learning methods of BLV people, teaching techniques and expertise of the teachers, familiarity with technologies, challenges of learning or teaching body movement on a broader level. Then from the followup interviews \textbf{(Section 5)}, we learned their individual challenges and motivations in accessing body movement. Those identified challenges were then discussed as separate focus groups \textbf{(Section 6)} with BLV participants and teachers to understand further and identify their priorities. We extracted 4 main themes (Techniques and Expertise of Teachers, In-person learning, Online learning, Body movement elements) and further 15 sub themes by conducting a thematic analysis to both interview and focus group study scripts. Finally, we present our findings and discuss them in the context of existing research \textbf{(Section 7)}. 
From the thematic analysis of the data, \textbf{the primary contribution of this paper is four key design challenges} that we encourage the assistive technologies community to consider when co-designing potential solutions to improving access to learning body movement for BLV people and supporting their teachers in providing an inclusive body movement education: (1) Representation of body movement to support verbal instructions, (2) Supporting feedback and improving kinesthetic awareness, (3) Supporting spatial and social interactive body movement learning and (4) Supporting accessible remote learning. 

%Further, we highlight the importance of supporting their teachers in providing an inclusive body movement education. 

%As further clarifications, 2) we aimed at BLV \textbf{adults} as ~\cite{brian2020motor} states that most studies related to motor skill development have considered children with little to no research on young and older adults, lacking the nuance of adult education needs, 3) we refer to participants who are blind or have low vision as BLV people, BLV adults, or BLV students interchangeably, 

\section{Related Work}

In this section, we first discuss the exploration of barriers to engagement in body movement for blind or low vision %(BLV) 
people and briefly explain the need for further research %referred to different contexts. 
We then describe current educational practices for students with vision impairments. Finally, we discuss technological innovations of different body movement access methods. %Further, we provide insights in each section where the gaps exist. 
Note, we use \textbf{`body movement'} instead of physical education or exercise not to limit our focus as there is a gap in the literature of less focus on creative physical activities such as dance. 

\subsection{Exploration of Body Movement Educational Needs of BLV People}
\label{sec:relatedwork}

Research on physical education and sports for BLV people~\cite{haegele2019, capella2007, holbrook2009, marmeleira2014, jaarsma2014} indicates that the participation of BLV people in body movement activities is less compared to sighted populations due to multiple barriers. Some barriers identified are travelling to physical education programs~\cite{Lee2014, jaarsma2014}, reduced opportunities for social interactions ~\cite{hastbacka2016, jaarsma2014, brunes2017}, safety concerns~\cite{brunes2017, india2021}, lack of inclusive educational programs~\cite{boswell2022} and not having high quality instructors~\cite{brunes2017}. For instance, Mycock and Molnár~\cite{mycock2020} provide knowledge on the lived experience of the first author as a coach transferring from the dominant coaching paradigm (the ableist model) to coaching blind football that requires additional attention to numerous challenges such as the significant gap in key skill sets, the additional care, negotiating infrastructural shortcomings, developing different coaching styles and strategies along with assuming a support role. Studies in physical activities of BLV people identify the use of software and computing devices as a positive factor and facilitator for the participation of BLV people~\cite{jaarsma2014, marmeleira2020}.  It is recommended for research in access technology~\cite{siu2019} to explore the problems and needs first,  increasing the likelihood of developing technology that addresses a valid user need~\cite{Rector2020bookchapter}. Few studies, however, have investigated the needs of BLV people when designing technology-driven solutions for body movement participation, either initiating an exploration study~\cite{rector2015, india2021} or brainstorming ideas~\cite{rector2018, watanabe2022}. %\begin{quote}``Recommendation 1: begin at the beginning (start with the problem, not the technology)...%When wanting to make an impact with assistive technology, research must always start with a problem of practice before identifying how technology will fit the problem. '' \end{quote}
\begin{figure*}[ht]
  \centering
  \includegraphics[scale=0.2]{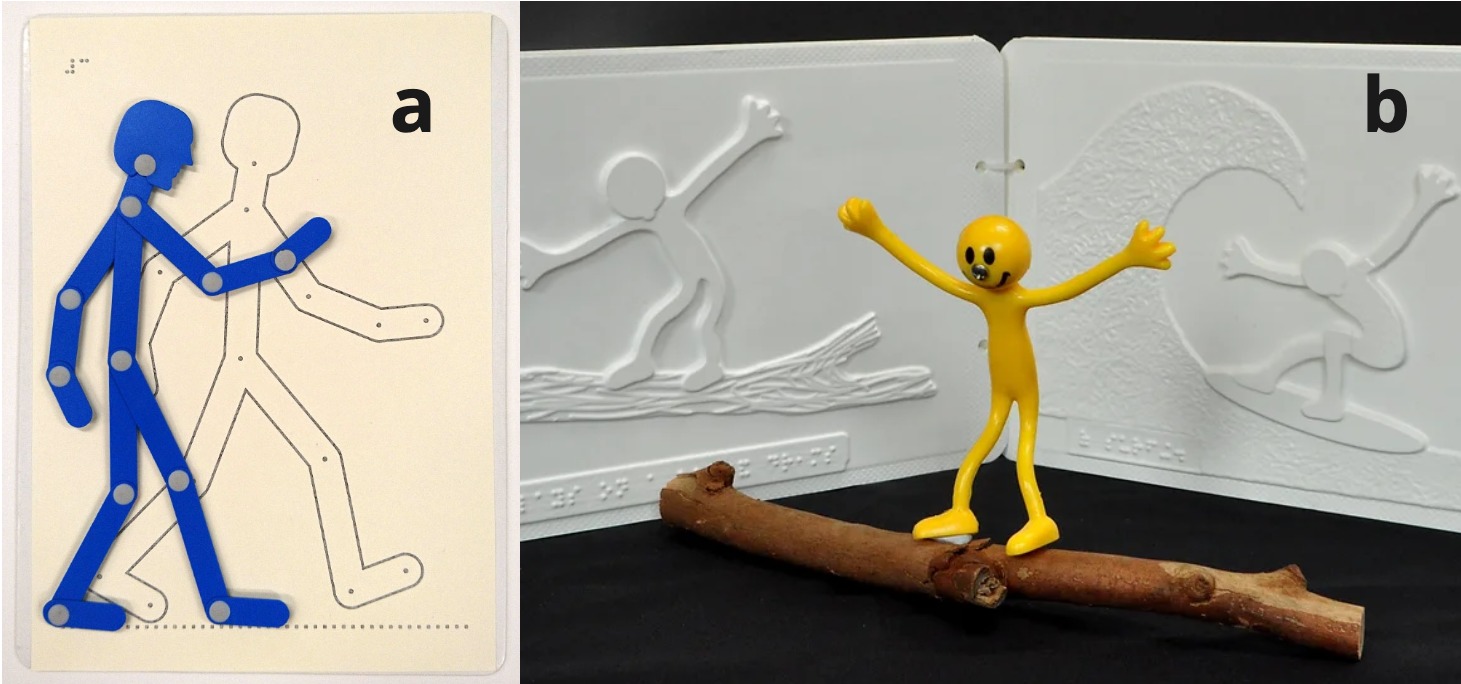}
  \caption{Examples of traditional approaches to conveying movement with movable parts: (a) Fleximan figure with movable parts (Picture credit: Boguslaw Marek, Hungry Fingers) (b) Manipulative Andy Dreams figures with a tactile illustration of the same pose behind it. (Picture credit: childsPly Vision by Claire Garrett).}
  \Description{As described in the caption. Image a Fleximan is a flattened stick figure with articulated joints that are manipulative to accompany a static tactile diagram. Image b Andy is a 13cm high figure that can be manipulated to match the pose of a same-sized tactile illustration of Andy, playing in the picture on the left side of each double-page spread.}
  \label{fig:representations}
\end{figure*}

\subsection{Body Movement Educational Practices for BLV People}

The general ways of transferring motor skills are using pictorials, verbal descriptions and mirroring the teacher ~\cite{adams1987, bandura1997}. Duggar~\cite{duggar1968} explains how body movement is taught in a visual nature, stating, ``Do it like this'' when it is meant to be developing a kinesthetic sense which involves bodily movement and awareness, by relying on our own body experiences to understand information. Seham and Yeo~\cite{seham2015} state the importance of not relying on visual cues when teaching blind people: ``When teaching the blind, however, the initial concept of each step, pose, position, or movement must be conveyed without reliance on visual cues''. A primary non-visual technique of teaching body movement to BLV people is verbal instructions ~\cite{seham2012, haegele2019}.  However, adapting the right terminology to suit BLV students requires investment in time and skill of the teachers ~\cite{zitomer2016}, which can be quite challenging.

Physical interactions or touch demonstrations are generally used as a complementary teaching technique with verbal descriptions when training BLV people in body movement ~\cite{paxton1993, oConnell2006, haegele2019}. \textbf{Physical interaction} can take the form of physical guidance or tactile modeling. \textbf{Physical guidance} is the use of physical assistance by an instructor to convey the feel, rhythm, and motion which includes physical placement of the student’s body part. In contrast, \textbf{tactile modeling} is the inspection by a student of a demonstrator by touch. Both of these methods have inherent benefits and challenges. Physical guidance can lead to misinterpretation and might not be the option for students who are sensitive to touch ~\cite{oConnell2006}. A similar problem exists with tactile modeling in which the personal space of the instructor or a peer is intruded upon and might not be the most appropriate technique ~\cite{esatbeyoglu2021}.

\subsection{Representations of Body Movement for BLV people}
\begin{figure}[ht]
  \centering
  \includegraphics[scale=0.30,angle=-70]{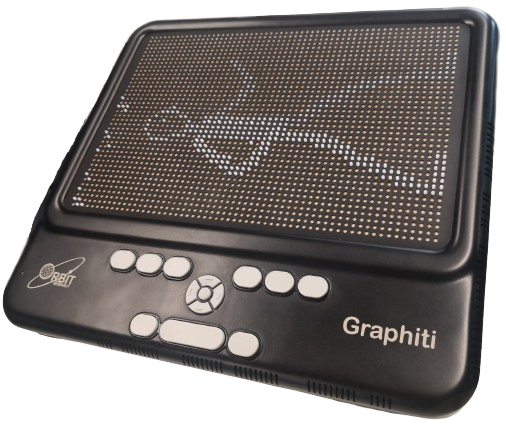}
  \caption{A refreshable graphics display, Graphiti displaying a Tai Chi human pose.}
  \Description{Graphiti is a refreshable display by Orbit Research. Left to the display of pins are a series of buttons arranged in 3 rows. The top row consists of 6 braille input buttons. The middle row consists of four directional buttons with a select button in the centre. The bottom row has two input buttons and a space bar. The Graphiti displays a simplified human figure posing in a Tai Chi sequence with a body 5 pixels wide and limbs 2 pixels wide.}
  \label{fig:taichi}
\end{figure}

Traditional methods for non-visual access to information include verbal description and raised line drawings. Audio description of events for people who are BLV is most commonly used for theatre, television and movies but also extends to sporting events~\cite{Fryer2016,Snyder2014}.   
Raised line drawings are recommended as the best way for people who are blind or have low vision to access and understand diagrams with spatial information~\cite{BANA2010, Landau2013, RoundTable2022}. However, guidelines for the design of raised line drawings assume that they will be based on static images ~\cite{BANA2010, Edman1992, RoundTable2022}. Another approach is the use of tactile graphics~\cite{kim_toward_2015} or models with movable parts, such as Fleximan by Hungry Fingers (Figure ~\ref{fig:representations}a, ~\cite{boguslaw2011}) and manipulative figures of Andy Dreams by (Figure ~\ref{fig:representations}b, Claire Garrett~\cite{claire2023}). Furthermore, some tools, such as refreshable graphics displays~\cite{kobayashi2021, holloway2022}, have been evaluated in representing body movement concepts (Figure \ref{fig:taichi}). For instance, Holloway and colleagues~\cite{holloway2022} compared the representation of body poses on a refreshable graphics display and an equivalent tactile graphic, finding that the graphic display was more successful. In addition, sound is commonly used to provide information about  positioning in space for sports, for instance, with bells inside balls ~\cite{haegele2019} for activities such as goalball and blind cricket. 

%Several studies have been conducted aiming to support BLV people's participation in body movement activities such as sports and exercise. 
\subsection{Technological Support for Body Movement Participation of BLV people}

Many accessibility technologies based studies have aimed to assist BLV people in body movement to improve spatial awareness without relying on sighted guides~\cite{al2016, ding2019, folmer2015, richardson2022, cooper2022, long2016, muehlbradt2017, oommen2018, padman2020, rector2018, sadasue2021} and to enhance performance ~\cite{miura2018, watanabe2022, miura2018b, junior2022, ramsay2020}. Miura, Watanabe and colleagues~\cite{miura2018, watanabe2022} implemented a training application for goalball players to predict the direction, height, and distance of an approaching ball using binaural sound. However, fewer have been designed to support learning or understanding how to perform a body movement activity ~\cite{rector2017, dias2019}. Dias and colleagues~\cite{dias2019} have developed a solution combining a web-based interface and a commercially available body tracking device, to provide the students with pre-recorded dance instructions and synthesized audio feedback for understanding body poses and directional cues. The need for understanding different actions in engaging body movement is a clear need which was also pointed out by Morelli and colleagues~\cite{morellivi-tennis2010} who experimented with sensory substitution of exergames such as the VI-bowling. They share that they verbally instructed how to play the games but found that ``some children did not know or understand how to swing a tennis racket'' while some have developed ``completely new ways to swing their racket''. This leads to the need for exploring what challenges exist when BLV people are learning body movement and correct techniques. 
%of body movement in blind sports such as goalball ~\cite{miura2018, watanabe2022}, blind hockey ~\cite{cooper2022}, sound table tennis ~\cite{miura2018b} and paralympic running ~\cite{junior2022, folmer2015}.  Some have been attempting to adapt sports such as such as skiing ~\cite{aggravi2016} and swimming ~\cite{muehlbradt2017, oommen2018, padman2020} through sensory substitutions. Several studies have involved  exercise-related activities such as aerobics [Malik 2021], track-based walking ~\cite{rector2018}, and exergames such as the Vi-Bowling ~\cite{morellivi-bowling2010}, Vi-Tennis ~\cite{morellivi-tennis2010} and yoga exercises ~\cite{rector2017}. While many studies have chosen the contexts of sports, and exercise, relatively little research ~\cite{dias2019} has addressed the accessibility of learning creative body movement, such as dance, for the BLV community from a technology support perspective. The majority of these studies have addressed enhancing performance and assisting needs, while a few have been designed to support learning or understanding a body movement activity ~\cite{rector2017, dias2019}. 

\begin{figure*}[ht]
  \centering
  \includegraphics[width=\linewidth]{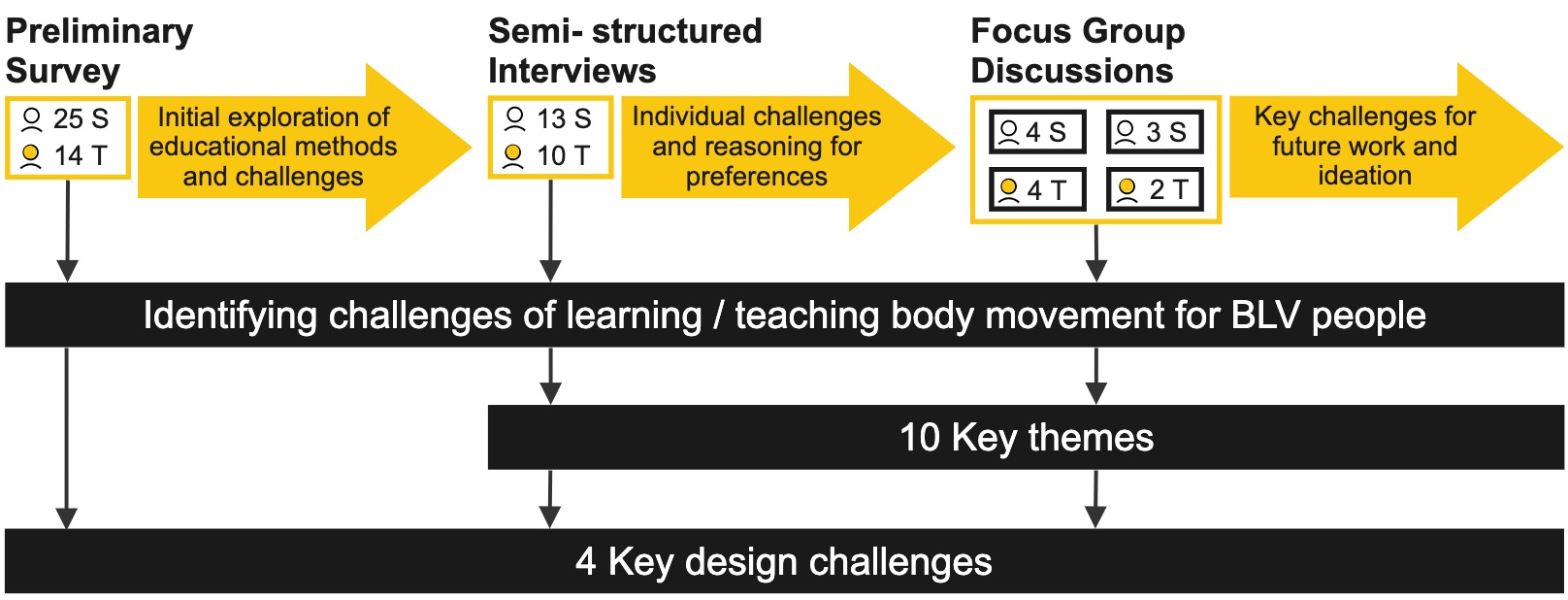}
  \caption{The methodology and the key findings of the process. Participant categories of BLV students are denoted by ‘S’ and teachers by ‘T’ in yellow boxes. Input for the next stage in yellow arrows. All three stages contribute to identifying challenges in learning and teaching body movement. Analysis from interviews and focus groups contributes to identifying ten key themes. Finally, all findings contribute to deriving four key design challenges.}
  \Description{The methodology and the key findings of the process as described in the caption. The overall process is provided as three separate sections entitled to each preliminary survey, semi-structured interviews and focus group discussions connected from one after the other sequentially. Under each section, the number of participants for each category of BLV students and teachers is listed. For the preliminary survey, it is 25 BLV students and 14 teachers. For the semi-structured interviews, it is 13 students and 10 teachers. For the focus group discussions, the number of participants is listed under 2 groups for each category. For BLV students, one group has 4 students and the other has 3 students. For teachers, one group has 4 teachers and the other has 2 teachers. Then there are connections between each study section indicating the input that has been carried forward to the next stage. Findings of an initial exploration of educational methods and challenges are input from the preliminary survey to semi-structured interviews. Findings of individual challenges and reasoning for preferences are input from semi-structured interviews to focus group discussions. Findings of key challenges for future work and ideation are input from focus group discussions to future undefined stages. All three sections contribute to identifying challenges in learning and teaching body movement. Analysis from interviews and focus groups contributes to identifying ten key themes. Finally, all findings contribute to deriving four key design challenges.}
  \label{fig:methodology}
\end{figure*} 

With the Covid-19 pandemic, there was a trend in supporting online education, including physical and mental health-focused activities such as exercise programs and yoga ~\cite{brinsley2021, brosnan2021, mcDonough2022}. Some support BLV people to adapt and engage virtually in different body movement categories such as sports ~\cite{kim2016sonic, morellivi-tennis2010}, exergames ~\cite{morellivi-bowling2010} and yoga exercises ~\cite{rector2017}. Both pre-pandemic ~\cite{miura2018, rector2017} and recent post-pandemic ~\cite{watanabe2022, dias2019} studies have attempted to support training or learning body movement remotely. Few have explored whether those solutions are practical and sustainable to use at home beyond a lab environment. Rector and colleagues~\cite{rector2017} designed an exergame aiming for long-term engagement rather than short-term use in a lab setting for blind yoga learners at home. However, it is unclear, and there is a need for further understanding of how BLV people learn body movement with the recent and rapid development of online content and remote learning tools.%, especially in body movement.
%For instance, [Rector2017] is the only reported and ``first in-home deployment study of an exergame for people who are blind or low vision that looked at long-term engagement rather than short-term use in a lab setting''.

Some studies have consulted teachers and instructors of the relevant sport or activity. However, few have collaborated from the initial design or exploration stages ~\cite{rector2017, dias2019, cooper2022}. For instance, Cooper and colleagues~\cite{cooper2022} investigated multiple electronic prototype designs uncovering necessary characteristics of an audible-based hockey puck through iterative design based on feedback from blind hockey players and their coaches. Little work has considered the interaction between the student and the teacher as a research aim~\cite{aggravi2016, malik2021}. Aggravi and colleagues~\cite{aggravi2016} designed a haptic assistive bracelet that allows ski instructors to provide directional guidance to their students. However, it is unclear what challenges are experienced by the teachers and how these designs can help them to educate BLV people. %Exploring and addressing challenges experienced by teachers of BLV students is under-explored and requires the attention of inclusive technologies researchers and designers towards this perspective. 

While prior research has explored various technological approaches to improve access to participation in body movement activities, it still remains unknown how BLV people and their teachers adopt these methods. Further, most of the body movement educational practices remain separate from the technologies research space, and there is less evidence of initial explorations conducted involving people with lived experiences in understanding their challenges. Thus, our study explores an in-depth understanding of the challenges and opportunities of educating BLV people, offering  insights into how technology can facilitate their teachers to deliver a better outcome.

\section{Methodology}

\subsection{Methodological Overview}

With the aim of addressing the research questions and considering the gaps in the literature, we conducted a series of data collection with (1) BLV adult students interested in learning body movement (`S'); and (2) teachers who have experience in teaching body movement to BLV people (`T') (Figure~\ref{fig:methodology}). We first conducted a short survey to get a broad insight, and then we conducted interviews and focus groups to obtain a deeper understanding of key issues. %To participate in our studies, both participant groups needed to be 1) 18 years or older and 2) be able to communicate in English. 

\subsection{Preliminary Survey}

\begin{table*}[ht]
\begin{tabular}{p{2.5cm}p{7.1cm}p{6.7cm}}
\toprule
\textbf{Attribute} & \textbf{BLV Students} & \textbf{Teachers} \\ \midrule
\textbf{Age} & 25 - 75 and above, Median: 45-54 & 26 - 75, Median: 45-54 \\
\textbf{Gender} & F (9), M (3), N (1) & F (5), M (4), N (1) \\
\textbf{Country} & Australia (9), USA (4) & Australia (8), Egypt (1), Netherlands (1) \\
\textbf{Blindness} & Totally blind (7), Legally blind (5), Low vision (1) & Legally blind (1) \\
\textbf{Onset} & Congenitally blind (6), later in life (7) & Later in life (1) \\
\textbf{Tech adoption} & Need support (3), Up-to-date (5), Early adopter (5) & Need support (3), Up-to-date (6), Early adopter (1)\\
\textbf{Body Movement} & Dance (8), Sports (9), Martial arts (8), Yoga (6), Circus (1) & Dance (6), Sports (4), Physical education (3), O\&M (1)\\ \bottomrule
\end{tabular}
\caption{Demographic information of interview participants. O\&M stands for Orientation and Mobility. Gender is listed by referring to female as `F', male as `M' and non-binary as `N'.}
\label{tab:overallInterviewParticipants}
\end{table*}

\subsubsection{Instrument}
The survey began with obtaining demographic information, including frequently used assistive technologies. We considered ‘body movement’ as recreational physical activities such as sports, yoga, exercise and dance while allowing participants to state a variety of physical activities they engage in. We then asked BLV students about their interests in body movement and methods of accessing body movement education. In the survey for teachers, we asked about their expertise in body movement and the methods they use in teaching body movement for BLV people. In both surveys, we asked them to list challenges they experienced when learning or teaching body movement.

\subsubsection{Recruitment and Participants}
The surveys were distributed online using Google Forms. They were tested for accessibility, and keyboard shortcuts were provided. The survey link was shared through an approved participant pool, social media, listservs~\footnote{Listservs are electronic mailing list systems that enable group communication and facilitate the distribution of information among subscribers with shared interests.} and organisations that educate BLV people on body movement. A total of 25 people responded to the BLV student's survey and a total of 14 to the teacher's survey (Figure~\ref{fig:methodology}). The median age range of the BLV students was reported to be 35-44, and for teachers, it was 45-54. The majority of BLV students (64\%) and teachers (78\%) were from Australia. Some BLV students (32\%) were from the United States, and one from Canada. The remaining teachers were from the United Kingdom, Egypt and Netherlands. A majority of the BLV students self-identified as female (68\%), 20\% as male and 8\% as non-binary. Similarly, the teachers mostly self-identified as female (57\%), 28\% as male and 14\% as non-binary. Of the BLV students, the blindness condition was self-reported as 44\% totally blind, 48\% as legally blind (not totally blind but qualifies for government aid~\cite{AIHW}), and two as low vision stating that they have ``peripheral vision loss'' and ``left eye cataract right eye no vision''. One of the teachers self-reported as legally blind, and they had acquired blindness before age 30. 44\% of the BLV students are congenitally blind. For the remaining students, the average onset of blindness was 16 years (SD = 14.5). 40\% of BLV students considered themselves either up-to-date with new technology or early adopters. Most teachers (64\%) considered themselves up-to-date with new technology. 

%(n(BLV students) = 16, n(teachers) = 11), with the remainder in the United States (n(BLV students) = 8), Canada (n(BLV students) = 1), the United Kingdom (n(teachers) = 1), Netherlands (n(teachers) = 1)  and Egypt (n(teachers) = 1). Seventeen BLV students self-identified as women, five as men and two as non-binary. Eight teachers self-identified as women, four as men and two as non-binary. Of the BLV students, 11 were totally blind, and 12 were legally blind, 2 as low vision. One of the teachers was legally blind, and they had acquired blindness before the age of 30. Of the BLV students, 11 were are blind or have had a low vision from birth; for the remaining 14 students, the average onset of blindness was 16 years (SD = 0.38). Ten BLV students each considered themselves up-to-date with new technology and early adopters. Most teachers (n=9) considered themselves up-to-date with new technology.

\subsubsection{Analysis}

As our intention was to gain high-level insight, we have provided a descriptive summary of the data collected from the surveys in section~\ref{sec:surveyresults}.  We summarised the responses provided as the interests of BLV students and the expertise of teachers. We then compared the current educational methods across the two participant groups: BLV students and teachers. Finally, we have briefly listed the challenges shared by both participant groups.

\subsection{Semi-structured Interviews}

\subsubsection{Study Procedure}

We conducted semi-structured interviews to expand the understanding gained from the surveys with BLV students and teachers. The interviews were intended to understand their individual experiences in learning or teaching body movement, including any reasoning behind the teaching method selections teachers adopt to provide access to body movement, what challenges are encountered, and the motivations of BLV people behind their preferences. Participants were asked if the Covid-19 impacted their learning or teaching body movement and their opinions on using technology-based methods. 
%Based on their survey responses, BLV participants were asked the specific reasons for their preferences of learning or participating in particular body movement activities. We also asked if there are any activities that they have never tried or wished to attempt and what stops them. BLV participants were also questioned on the reasoning behind their preferences in modes of accessing lessons. For instance, particular reasons as to why they prefer to attend a BLV-specific body movement class or otherwise. We asked about their learning journey in a class, such as how they would reach the class, set themselves up with others, what ways their teachers have taught them about different body movement elements, and how they perceived and felt about those instructions. We also asked what other ways they learn body movement besides attending classes.
We questioned the teachers on their experiences and reasoning regarding the different modes of conducting classes for BLV people. For instance, how they feel about conducting BLV-specific classes vs inclusive classes and in-person vs online classes. Then teachers were asked what techniques and tools they use and in what situations.

\subsubsection{Recruitment and Participants}

Thirteen BLV students and ten teachers who consented to join follow-up discussions were recruited for the interviews (Figure~\ref{fig:methodology}). All interviews were conducted by the same researcher, scheduled for the convenience of the participants. Refer to Table \ref{tab:blvParticipants} and Table \ref{tab:teachers} for the individual demographic information of the interview participants, which also includes the expertise of teachers. %These tables also provide BLV students' body movement interests and teachers' expertise, including the self-reported number of years of experience and the number of BLV students they have taught. 
The median age range of both BLV students and teachers was 45-54 (Table ~\ref{tab:overallInterviewParticipants}). The vast majority of BLV students (69\%) and teachers (80\%) were from Australia. %, with the remainder in the United States (n(BLV students) = 4), Netherlands (n(teachers) = 1)  and Egypt (n(teachers) = 1). 
A majority of BLV students self-identified as female (69\%), 23\% as male and 7\% as non-binary. Half of the teachers self-identified as female, 40\% as male and 10\% as non-binary. Of the BLV students, the blindness condition was self-reported as 53\% totally blind, 38\% as legally blind,
and 7\% as low vision. 46\% of the BLV students are congenitally blind, and the average onset of blindness was 20 years (SD = 16.61) for the remaining students. One BLV student has been legally blind since birth and has become totally blind when eight. One of the teachers self-reported as legally blind, and they had acquired blindness before age 30.

\subsubsection{Analysis}
\label{sec:interviewanalysis}
The interviews were conducted online or by phone %at the participant’s convenience 
for around 45-60 minutes. All interviews were audio-recorded and transcribed. The transcripts of all participants were analyzed together and subjected to a thematic analysis based on a qualitative approach using both inductive and deductive coding in several rounds~\cite{saldana2021}. The transcripts were inductively coded in the first coding round to identify challenges, learning methods and teaching techniques. Deductively, furthermore, codes were identified such as `physical contact', `support from another', and `safety considerations'. Three researchers discussed and grouped the identified challenges under four key themes of body movement elements. Then those themes and challenges were subjected to further discussion with focus groups of the same participant pool since they were frequently mentioned in the interviews and required further understanding.

Next, two independent researchers validated codes in the initial round by randomly sampling transcripts. They coded five random transcripts (three transcripts of BLV students and two transcripts of teachers) and discussed any discrepancies. Based on those discussions, a final codebook was formed, and one researcher recoded all interview transcripts. A final validation round was conducted by another researcher examining a subset of the transcripts. They confirmed all coded parts were correct. %and identified three new potential codes in the subset. However, after discussion, it was determined that existing codes already represented them. 

\subsection{Focus Groups}
The focus group discussions with BLV students and teachers aimed to confirm further and understand the challenges identified from the interviews. In addition, we intended to identify their priorities among those challenges. 
\begin{table*}[ht]
\begin{tabular}{p{2.2cm}p{3cm}p{3cm}p{3cm}p{3cm}}
\toprule
Attribute & Students FG1 & Students FG2 & Teachers FG1 & Teachers FG2 \\ \midrule
Gender & F (3), M (1) & F (1), M (1), N (1) & F (2), M (2) & F (1), N (1) \\ \midrule
Blindness & Totally blind (1), Legally bind (2), Low vision (1) & Totally blind (2), Legally bind (1) & Legally blind (1) & - \\ \midrule
Onset & Congenitally blind (1), later in life (3) & Congenitally blind (3) & Later in life (1) & - \\ \midrule
Body Movement & Dance (3), Sports (4), Martial arts (3), Yoga (3) & Dance (1), Sports (2), Martial arts (1), Yoga (1),  Circus (1) & Dance (2), Sports (2), Physical education (1),  O\&M (1) & Dance (1), Sports (1), Physical education (1)\\ \bottomrule
\end{tabular}
    \caption{Demographic information of focus group (FG) participants. O\&M stands for Orientation and Mobility. Gender is listed by referring to female as `F', male as `M' and non-binary as `N'. All participants were from Australia.}
    \label{tab:overallfocusgroups}
\end{table*}
\subsubsection{Study Procedure}
One researcher facilitated all focus groups, and another researcher observed the process. The focus group discussions were conducted online for around 90-120 minutes. 

Each focus group comprised two discussion rounds. In the first round, participants were provided with a summary of the challenges discovered from the interviews (section~\ref{sec:interviewanalysis}) with a focus on body movement elements (body awareness, spatial awareness, etc.). Participants were asked to share their opinion on those challenges, select what they believed was important to be addressed first and if there was anything further to be considered. Finally, in the second round, the facilitator summarised what was discussed, confirming the notes taken and asked the participants again to express what should be prioritised based on the overall session. %Three researchers brainstormed on the agenda and the topics of the discussions. 

\subsubsection{Recruitment and Participants}

The interview participants who provided consent and who were able to join in a common time frame were invited for collaborative focus group discussions. One participant who could not join the interview also joined the focus groups responding to a previous invitation. %Seven BLV students and six teachers joined in this continuing discussion series (Figure~\ref{fig:methodology}). 
The focus group discussions were conducted separately with the BLV students and teachers to encourage open discussion and avoid potential power dynamics. Two focus group discussions were conducted with BLV students, with one having four participants and another with three participants (Table \ref{tab:overallfocusgroups}).  Similarly, for teachers, two focus group discussions were conducted, with one having four participants and another with two participants. %The median age range of both BLV students and teachers is 45-54. 
%All participants resided in Australia. %Four BLV students self-identified as women, two as men and one as non-binary. Three teachers self-identified as women, two as men and one as non-binary. 
%Three of the BLV students were totally blind, and four were legally blind. One of the teachers was legally blind. %, and they had acquired blindness before the age of 30. Of the BLV students, four are blind or have had a low vision from birth. 

\subsubsection{Analysis}

All focus group discussions were audio-recorded and transcribed. Then the four scripts were coded using the same codebook derived from the interview analysis. However, a new category code was introduced to capture responses relating to the prioritisation of challenges, which had not emerged earlier and was a key element of the focus groups. 

\section{Results}

\subsection{Overview}

In this section, we first present the survey findings, including the intentions set for the interviews. We then report on emergent themes from both interviews and focus group studies. These are presented together owing to two reasons: first, the focus groups predominantly expanded upon topics from the interviews; and second, we observed repetition of some experiences shared by the participants. However, findings from the focus groups confirmed many insights from the interviews and identified key priorities, which are noted as appropriate.

\subsection{Survey}
\label{sec:surveyresults}

\subsubsection {Interests and Expertise}
Interestingly, the majority of BLV student respondents (n=20) were interested in learning dance. Other interests were sports (n=16), martial arts (n=11), yoga (n=9) and other types of activities (n=11), including tai chi, circus, bowling, and kayaking. 
%Here we identified yoga and tai chi separately from martial arts as there are arguments around them~\cite{buschbacher_martial_1999,channon_im_2015}. 
One respondent specifically listed dance as an activity they stopped learning. Another respondent who did not select dance stated in the survey about dance: \begin{quote} ``I feel hesitant to participate in case I embarrass myself by moving the wrong way'' %It makes me feel very `blind'. This is also why I never dance! ''
\end{quote} 

The teachers who responded to the survey mainly work as dance teachers (n=8) and experts in sports (n=6), while some work as physical education teachers (n=3) and a few are experts in yoga, and orientation and mobility. Most teachers have experience educating BLV children and adults (n=10), while four teachers have only taught adults, and two teachers have taught only children. 

\subsubsection {Current Learning or Teaching Methods}
\label{sec:surveylearningteachingmethods}
(Table \ref{tab:accessmethods}) summarises the learning methods used by BLV students, along with the methods used by teachers. All BLV students responded (n=25) they learn body movement by verbal instructions provided by a teacher or someone they know. The next most frequently used learning method was physical guidance by an instructor (n=18). The remaining methods of access were tactile modelling (n=9), videos (n=13), and written descriptions (n=3). 

All teachers (n=14) verbally instruct their BLV students, while most of them use physical guidance and tactile modelling (n=12), except for two teachers who conduct online sessions. Sound-based tools such as whistles, drums, and clapping are being equipped by some teachers to guide their BLV students (n=10). Some of them mentioned how they used sound-based techniques,  like ``African drums to cue movement'', and %“Actually I don't use a clicker or whistle, but do use 
``clapping to guide students if, for instance,  they are running across the room''. Four teachers share educational content through instructional videos and use 3D models. Three teachers use instructional images. One teacher commented about using images, ``Images not so useful in a group setting either and would have to be shown up very close''.

\begin{table}[ht]
\begin{tabular}{@{}lrr|rr@{}}
\toprule
\begin{tabular}[c]{@{}l@{}}Suggested learning/ \\ teaching method\end{tabular} & \multicolumn{1}{l}{\begin{tabular}[c]{@{}l@{}}Teachers\\ (n=14)\end{tabular}} & \multicolumn{1}{l|}{\begin{tabular}[c]{@{}l@{}}BLV\\ (n=25)\end{tabular}} & \multicolumn{1}{l}{\begin{tabular}[c]{@{}l@{}}T\\ (n=11)\end{tabular}} & \multicolumn{1}{l}{\begin{tabular}[c]{@{}l@{}}L\\ (n=14)\end{tabular}} \\ \midrule
Verbal instructions & 14 & 25 & 11 & 14\\
Physical guidance & 12 & 18 & 9 & 9\\
Tactile Modelling & 12 & 9 & 5 & 4 \\
Sound-based tools & 10 & 0 & 0 & 0 \\
Videos & 4 & 13 & 7 & 6 \\
3D models & 4 & 0 & 0 & 0 \\
Instructional images & 3 & 0 & 0 & 0 \\
Written descriptions & - & 3 & 1 & 2 \\ \bottomrule
\end{tabular}
\caption{Body movement educational methods used by BLV students and teachers.  `T' represents totally blind students, `L' represents legally blind and low vision students}
  \label{tab:accessmethods}
\end{table}

Although some teachers use sound-based tools (n=10), 3D models (n=4), and instructional images (n=3), none of the BLV students experienced such tools. Three BLV students shared the use of written descriptions in the survey, but none of the teachers mentioned their use of them. Half of the blind (n=7) and low vision (n=6) students make use of videos to learn body movement. However, only four teachers stated the use of videos as educational materials.
\begin{table*}[ht]
\begin{tabular}{@{}lc@{}}
\toprule
Challenge & \begin{tabular}[c]{@{}c@{}}BLV students \\ (n=25)\end{tabular} \\ \midrule
I do not get feedback on whether I am doing the pose correctly & 18 \\
Instructions alone are not enough for understanding human movements & 15 \\
I do not have access to learning materials such as videos in accessible formats & 15 \\
I cannot learn with the speed of the instructor & 15 \\
I cannot follow floor markings or where to move in the space & 14 \\
I cannot follow objects used in the activity such as a ball when playing football & 13 \\
Descriptions are not provided in learning materials & 12 \\
Verbal instruction does not provide feedback on emotions, body language and other aspects & 10 \\
I cannot follow where to move in the space with other students & 10 \\
It is difficult to learn without tangible materials to touch & 9 \\
I do not like physical touching & 2 \\ \bottomrule
\end{tabular}
\caption{Challenges of learning body movement for BLV students}
  \label{tab:challenges}
\end{table*}
\subsubsection {Challenges of Learning Body Movement} Table \ref{tab:challenges} summarises the primary challenges in learning body movement, as identified by BLV students. The majority of the BLV student respondents (n=18) find it challenging to learn body movement since they don’t receive feedback. Many BLV students (n=15) believe instructions alone are insufficient to understand body movement. Some BLV students find difficulties in following floor markings and boundaries (n=14), following objects played with (n=13) and moving with others (n=10). Regarding learning materials in general, such as videos, 15 students responded that they do not have access to them in accessible formats, while 12 responded to not having descriptions of those learning materials, and nine students found it difficult to learn without tangible materials to touch. Two students do not prefer physical contact. One student interested in martial arts elaborated in the survey how physical contact is important for a congenitally blind person and the impact of having more people to support:  
 \begin{quote} ``Being totally blind from birth, I have not had the experience of seeing others do a physical activity, so even mentioning, for example, that the movement is similar to holding a baseball bat doesn’t mean much to me. The very best way to learn is to have the instructor make the move on me and then for me to practice the move with two others so that there is someone to observe my movements on my partner'' \end{quote}

 When BLV students were provided with a list of challenges in the survey, teachers were provided with a free text field input because the teacher's perspective has received less prior exploration (section~\ref{sec:relatedwork}). Some of the challenges listed by teachers were around verbal instructions such as ``Difficult to communicate subtleties of movement/style'', ``It took me a while to remember I had to say everything, so much more talking than in an ordinary class where you... rely on what the students can see to fill in the gaps'', or ``Sometimes I struggle with what words to use to describe something to someone who has never had the vision to see movement''. One respondent mentioned that BLV people's lack of body awareness is ``problems with body plan, body awareness and body idea''. Another respondent shared, ``Poor spatial awareness,  lack of body concepts, poor balance, fear of moving in space''. 
 
 %A teacher suggests that it can be challenging to use some tools, such as sound-based and creating 3D modals, ``tools such as whistles will be a distraction / sensory issue for other students in the class. 3D models are not something we have the budget or time to work with during a class. This could work for 1:1 lessons''.

\subsubsection{Summary}

BLV students who responded to the survey showed interest in learning a variety of body movements. However, there were activities reported that they have stopped or wish to learn in future, such as dance which needs further understanding as to why. There was a mixture of experiences among teachers in educating BLV children and adults that could be insightful if there was any difference in educating them. Instructions and physical guidance were the most used methods. However, there were some indications from BLV students on the insufficiency of instructions for understanding body movement and a few not preferring physical contact. Further, some teachers indicated difficulties in communicating movement. Both BLV students and teachers indicated problems in spatial and body awareness. There are some mismatches in some tools used by teachers and the exposure by the BLV students to those tools. 
%Given the limitations of the survey nature and the insufficient number of participants, it is difficult to deduce correlations or justify them. 
Thus, we conducted interviews to understand the reasoning behind the survey responses.

\subsection{Interviews and Focus Groups}

Results from both interviews and focus group discussions are provided here. Ten themes emerged from the analysis of the transcripts of both studies. Table \ref{tab:themes}, presented in the Appendix, provides an overview of themes that emerged in both or either of the studies. Overall the themes that emerged are related to practices in teaching BLV people, challenges, motivations and body movement elements. However, owing to the inherent interrelatedness of these themes, we have opted to present them as distinct entities, rather than grouping them under overarching categories. Please note that the participants are referenced using the prefix to represent the student (`S') or teacher (`T'), the number to represent the participant ID and the suffix to represent the study (`I' for the interview or `F' for focus groups). For example, S3I refers to a BLV student who participated in the interviews and S3F refers to the same student but a statement made in the focus groups.

\subsubsection{Verbal Instructions}
\label{sec:verbalinstructions}

All BLV students expect verbal instructions to be detailed and specific. S9I explained, ``The teachers don't say I'm standing at a 45-degree angle with my feet a foot apart. They don't say that stuff''. Some BLV students (S2I, S3I, S4I, S14F) prefer instructions supported with metaphors, such as ``bend your arms and wave them like you're a bird flapping their wings" (S3I) or ``move your body like a blade of grass, or a tree waving in the wind" (S14F). Three teachers mentioned that they use metaphors (T1I, T4I, T5I). However, three BLV students raised the issue that congenitally blind people might not understand or relate to the visual aspect of such metaphors (S5I, S6I, S7I). S5I shared, ``They asked for a movement. For what movement will you do for the colour yellow ... I've never seen yellow. I have no idea''.

From the teachers' perspective, some (T1I, T5I, T9I) found it difficult to explain the details and subtleties of body movement, which they could conveniently demonstrate visually. T1I recalls one of their experiences: 

\begin{quote} ``I tried to write down the movement and I remember looking back on it, I could not make sense of it. The main reason is, there are too many variables. In human movement, it's very complex... 
%When you sort of express something with your body there is so many little different things that is not just about where your arm is in space, it's about your weight, your attention, or maybe you are doing a movement like this (shows a hand movement flipping the palm). How are you gonna explain this? In words? Yes, you could explain that and then more things are going on like this, it's just complex. 
Words don’t do it justice.'' \end{quote}

One teacher (T5I) tried to empathise by using different techniques of their own such as closed eyes instructions. S11I shared their experience of learning from such a teacher.
%\begin{quote} ``She actually did the session with her eyes closed, and really thought about how the language that she could use to describe those movements. And how that would work for someone who can't see.'' \end{quote}
However, a low vision teacher T9I expressed that even if they have the empathy of having a blindness condition, she struggles in instructing her BLV students.
\begin{quote} ``I can understand like, what they're thinking or what they're going through, and then sometimes it's not helpful. I still struggle, because people's vision is different.'' \end{quote}

S11I stated that vocabulary is another contributing factor to the complexity of verbal instructions. They stated that the use of different terms to refer to the same or similar body poses and movements by different teachers is a challenge for BLV students. Further, they propose if there was a tactile dictionary of poses, it would be beneficial for cross-disciplines.

It was stated by several BLV participants (S1I, S2I, S4I, S7I) that they appreciate the understanding of how a body movement or a pose should feel and are not limited to how it should visually look like. S2I stated, ``describing how something should feel, instead of how something should look, is the best way to go''.

Two BLV participants (S2I, S3I) pointed out the effort required to concentrate on listening to the instructions. S3I stated ``most people get that double bit of information, the hearing and the scene. So I have to be dependent on the hearing''. Similarly, S2I shared ``I felt like I really had to concentrate and then listen to everything the teacher was saying''. Further adding to the focus on one medium of information (hearing instructions), T8I explained the need for additional modalities to support verbal instructions compared to sighted students who get double information based on verbal instructions and visual demonstrations,
\begin{quote} ``...if it was perfect, then that's all we would ever give our sighted students [who] always rely on visual and verbal. So we can't expect our blind students to just rely on verbal they need something else they need something tactile, and or audio feedback, of how it sounds when they're doing the movement.'' \end{quote}
Four other teachers (T3I, T5I, T9I, T10I) and two BLV students (S9I, S11I) expressed similar opinions.

\subsubsection{Physical Contact and Guidance}
\label{sec:physicalcontact}

According to many teachers (T2I, T3I, T4I, T5I, T6I, T7I T8I, T9I, T10I) and BLV students (S2I, S5I, S6I, S8I, S9I, S10I, S11I), teaching a particular body movement supported with physical contact-based instructions was expressed as a beneficial technique. S8I explained how his martial arts instructor has been helpful through a hands-on approach. Four teachers indicated how physical contact is supportive in teaching dance (T4I, T5I) and is inherent in some dance styles (T3I, T10I). %\begin{quote} ``we're teaching dance through body contact, as well as verbal, but we're really more relying on body contact, rather than verbal. And of course, that is how social dancing happens for traditional dancers anyway, that the leader will lead his partner through his weight transfers.'' \end{quote} %\begin{quote} ``...physical touch is something that all dance teachers find so incredibly helpful'' \end{quote} Supporting this, T5I said \begin{quote} ``...I would hold them, and I'd do it with them. And then they'd sort of feel this sort of slight little down and up down feeling that, so they get the rhythm of it and get the feeling and then gradually, we could sort of refine what they were doing '' \end{quote} 
 %, \begin{quote} ``So he was very hands-on with me, you know, grabbing my hand and showing me this is what you need to do.'' \end{quote}
%, you know, moving my legs a certain way with his hands if you need to do so I can get a feel for how to stand and so on. So he was really good in that respect}. 
From a sports and physical education perspective, T8I explained that teaching a particular body movement for the first time is effective through the support of physical guidance and  
%\begin{quote} ``verbal instruction needs to be quite accurate descriptive language but also the physical body mapping is also very beneficial to a student picking up a new move '' \end{quote}
%IS11 highlighted how physical guidance had less ambiguity than verbal description:
%\begin{quote} ``I think it'll say with description, there's so many variables where it can go wrong really quickly, like, you know, you only have to misuse one word, and someone can be doing it completely wrong. Whereas, you know, when it's being you being physically shown is a lot less room for error'' \end{quote}
T6I commented that physical contact is beneficial in providing feedback, \begin{quote} ``Professionally, it's a fantastic tool, because it can give you an instant way of getting a student feedback and giving them information about what the movement looks like, and how a complex movement might look and how much force to put into something, how fast to do something.'' \end{quote}

%IS11 being someone interested in dance but has given up stated,
%\begin{quote} ``I've never really got ballet. But just never I just never understood it. And I've tried to go and see some audio described. But you can't really dance things. And I just, I just, it's not a thing. Moving to music is unethical. Yeah, I can do a really big, I think it's that creativity thing that comes along with dance or been struggling with a bit not have to make it up'' \end{quote}

%IS11 further stated how not being able to watch a dance they did not know how they could move their body, \begin{quote} ``for me, never being able to sort of watch dance or things like that, I just didn't even have an idea about some of the ways your body could move, or the ways like that you could move your arms through space, for example. And so by having somebody else move your body for you, I think it was really interesting to feel the different ways that they would move your, your body through space, which might not have occurred to me.'' \end{quote}

There are, however, several factors contributing to the avoidance of physical contact in teaching body movement. Some BLV students do not prefer physical contact (S11I, S13I) and some have different preferences depending on the person and how it takes place (S2I, S3I, S4I, S7I). It matters to them whether they are comfortable with the teacher.
%as S7I stated \begin{quote} ``If I'm comfortable with the instructor, and there is a certain physical movement, that will help or I need to do, then I'm okay with that contact.'' \end{quote}
Another BLV student (S11I) preferred to bring a support worker or a friend who they are familiar with, \begin{quote} ``I took one of my really good friends along, she was my support worker. I was really comfortable with her life, so she could touch me if she needed to. I'm a little bit less confident having other people touch me. I would probably try to avoid it.'' \end{quote} 

Another preference factor is how physical contact takes place. For instance, whether it should be light touch or a strong one.  IS2 stated it might be acceptable for a light touch but T4I stated that they learned the touch should be firm. %\begin{quote} ``you just lightly touching them and giving like a light little push rather than grabbing and yanking '' \end{quote}
%However, T4I stated that they learned that the touch should be firm, \begin{quote} ``firm pressure is comfortable than light pressure. So if you think about when you're being tickled, it doesn't people, you if it's, it may take those lines. So it's all comfortable for somebody to be touched with a bit of shock, rather than sort of a light. So that was something that I learned. And it makes much sense. Because it's kind of the opposite of what you would do instinctively. '' \end{quote}
Consent plays a major role in physical interactions and it has an impact on body movement teaching.
%as well as participants pointed out, \begin{quote} ``If somebody just comes over and grabs me, I'm gonna react to that poorly. However, if somebody says, May I move your hand in this fashion, or put your hands if I can put my hands on top of theirs and watch what they're doing? So, it but it does have to be that consent. And if I don't feel comfortable with that person who's teaching? I'm not gonna give it.'' \end{quote} (S7I). 
Requesting consent is subjective to the person's preferences. During one of the focus group discussions, this was debated as one might need a consent request just once as S14F mentioned,``I'm more than happy for you to come in touch me and guide me in any way. You don't even have to tap me to do it. Just do it''. However, another participant in the same focus group (S3F) expected consent-seeking each time the teacher approaches them within a class, ``I'd be different...I do want you to tell me in advance... each time I like them to tell me they're gonna do it''. 
%Teachers have been respectful in terms of consent as assured by some participants such as IS11, "I've found that the instructors that I've been working with a very, very respectful, they will always ask before they touch my body, which is just fantastic". However, from a teacher's point of view this could impose a challenge with understanding the consent needs of every specific student as mentioned by IT7, "When they understand, okay, it's necessary. Now he wants to help me now or is for safety. That's okay. But often you have to say that you have to explain that every time".
%Furthermore, in terms of consent, FS1 expressed about people being cautious in physical contact and that it is not only a responsibility of teacher but the BLV student's as well to communicate their needs.
%\begin{quote} ``People are extremely nervous. And I think with political correctness, people are even more concerned about getting it wrong than ever...it's a bit of a two way street, you we can't put it all on the teachers that we have got a responsibility in naming really clearly what our needs.'' \end{quote}
Another factor for consideration in physical guidance is the health impact caused by the pandemic (S2I, S3I, S10I, T2I). %There is an increased reluctance for people to engage physically for the fear of contagious diseases as S2I mentioned, %\begin{quote} ``the teachers were quite good at doing adjustments but they weren't touching because of Covid.'' \end{quote}

%Teachers have also experienced BLV children and young adults being unaware of appropriate social norms on touching resulting inconveniences to teachers. IT8 shared that \begin{quote} ``I'm so used to having used my body to teach kids in that way '' \end{quote} IT2 expressed the multiple factors impacting the use of physical contact in teaching body movement, especially how cultural norms. \begin{quote} ``I feel sad about the loss of natural touch because culturally we have become much more nervous about touching people not just with Covid but sexualised touch, gender based stuff.'' \end{quote}

%One dance teacher (S10I) pointed out that some activities have inherent physical contact. 

\subsubsection{Tools Supporting Teaching} 
\label{sec:tools}

Some teachers are equipped with tools in addition to verbal instructions and physical guidance. Three teachers (T3I, T6I, T8I) mentioned the use of figures such as dolls and wooden human models. T8I described the use of such figures, \begin{quote} ``let's get my 3D model. They get their hands on the model. So this is where your foot was. And this is what your foot did. So that's one way they can get feedback is from me, recreating their movement through a tactile model.'' \end{quote}

%Further, IT3 described how the small size of those figures can be helpful in providing an overall idea of the body movement, \begin{quote} ``so imagine that your eyes is your hands. Okay, so you will see what you touch. So imagine, if you're touching something big, imagine you're touching a big statue. So you need to put a lot of bits and pieces together to realise the entire picture. So if I, when I am making small fingers, and small dolls, she could feel with one hand the entire shape of the person.'' \end{quote}

A similar idea was proposed by S13I on the use of flexible 3D structures that can bend well similar to the human body , ``I'm wondering whether what really would help would be rubber man, sort of figures that you could put into the position and so that they can be felt and then replicated''. 
%You'd also need something that could kind of bend in the middle as well. Because sometimes, you know, if you're doing something, you want to be able to bend in the middle, or you want to be able to bend at the hips, or, you know, wherever you are, you want to be able to bend sideways or whatever.'' 
%\end{quote}

Some teachers use sound-based tools such as balls with sound and squeaky toys (T7I, T8I, T9I). Recording the sound of movement to let BLV students listen is another strategy used by T8I to provide feedback. \begin{quote} ``Every very low vision or blind student who really relies on sound will have memorised that sound. They have to be good at that because  that's one of the ways that they make sense of the world is by understanding sound...when you're kicking your legs for a flutter kick in the pool, a good flutter kick, sounds a certain way...It's a video for those students, it's their way of playing back things.'' \end{quote}

Some teachers use wall-to-wall chords (T6I, T7I) and tactile diagrams (T8I) to indicate boundaries, floor patterns and, in sports, team members' positions. T8I explained that the game is halted in some situations to explain to the players which is challenging and impacts the flow of the game. %\begin{quote} ``We've always done this to provide tactile diagrams, either, you know, thermoform ones or ones made with raised line drawing paper %~\cite{wikki}...
%or ones with Wikki Stix or a little wax string, you can stick down onto paper... 
%We sometimes will stop a game and say, let's come back and look. These are where these players were... That's really difficult, kind of kills the moment.'' \end{quote}

\subsubsection{Community Knowledge Sharing and Practice}

Some professional bodies exist to support teachers to cater to BLV students' needs, such as in improving orientation and mobility and in contexts such as schools as shared by some teachers (T2I, T6I, T7I and T8I). T6I suggested that it would be beneficial if there were consultants who could support teachers who have BLV students in their educational programs. %\begin{quote} ``you can run professional development for teachers of the visually impaired but it doesn't always translate to good outcomes. I think what they could benefit from is a consultant that goes out and works with them in their actual schools to improve the delivery of phys ed, these students because one-day workshops are good stuff, but it's not enough'' \end{quote}
In the focus groups, it was observed that there is a need for community sharing on how they can accommodate BLV students. For instance, T5F explained their novice experiences in teaching BLV people, \begin{quote} ``%it's just nice to be able to talk to somebody who understands what you're doing. Because 
I've never spoken to anybody who's ever taught anybody blind...it's just good to have somebody who's, you know, on the same wavelength and understands what you are doing'' \end{quote} To which, then T8F responded that they could connect them with others who have similar experiences. %, ``If you're interested, I can put you in touch with all the right people...''. 
T6F and T9F appreciated and pointed out the importance of understanding the perspectives of different body movement experts. %\begin{quote} `` as both of you the different perspectives, also gives me some new perspectives to consider as well. Useful.'' \end{quote}
S6F shared that they teach and share their experiences with other blind people. 
%\begin{quote} ``I can now explain things better to other people...to a beginning student, I can explain it in such a way that they get it quicker than I got it.'' \end{quote}

\subsubsection{Spatial and Social Awareness}
\label{sec:spatialsocial}

Both BLV students and teachers raised challenges specific to spatial awareness. One that emerged several times is how BLV students find it challenging to position themselves in the class space with respect to others (S1I, S2I, S5I, S7I, S8I, S10I, S11I). For instance, S5I stated that one reason for bringing in a support worker is to ensure that others in the class are not being affected by them.
%, ``I don't know my location compared to other people most of the time...% That's quite difficult. Often, I'll have a support worker who can make sure I don't kill anybody''.
Six participants echoed the same challenge in three focus groups (S6F, S11F, S13F, T2F, T5F, T8F). However, two teachers shared their experiences of addressing it. An instructor with expertise in team sports (T9F) explained a strategy on how team players act as sound beacons supporting each other to locate themselves. T2F termed it as `beacon wayfinding' in Orientation and Mobility terms. %T9F stated the strategy as, \begin{quote} ``calling out to them working out where they are'' \end{quote} 
%Issues in spatial awareness were discussed around a few concepts in interviews and focus groups. 
T8F discussed about two types of movement skills in sports, open and closed, and the impact of environmental awareness: \begin{quote} ``In sport, we talk a lot about closed skills versus open skills. Closed skill is a skill that basically you're trying to do to absolute perfection because the skill never changes. 
%You think about people who swim freestyle in the Olympics, the pool length never changes, freestyle never changes, you never have to adapt it for someone coming into your lane or you know, you don't have to try to pass somebody, it's you literally just swim the same stroke. 
Whereas when you think about an open skill, that's a skill you must modify, depending on the environment and other players. 
%So you know, when you're dribbling a soccer ball, and yes, you might want to be dribbling down the field to score a goal. But suddenly, a player comes in your path. So what do you do, you have to do something differently, because environment has changed. 
The impact of that environment on learning is really it's not a small one''%; I think people often forget how important the environment is'' 
\end{quote}  

%FT2 pointed out another similar concept. They mentioned that whether the movement is planned or incidental has an impact on spatial awareness. It was also pointed out that Orientation and Mobility learning is universal and that it impacts body movement in general, \begin{quote} ``there is two ways of doing it, one is intentional route travel, going from point A to point B, there is no exactly one way to do it, there might be several paths. And that is different to incidental mobility. Incidental mobility happens in the context of everything else you do, you roll over when you feel wiggly, you hop out of bed to go to the toilet. When they get out of the seat and moving around having to find other ways than vision to know where their reference points are, where to go, how to stay orientated in the room, in the task, to get a sense of direction move efficiently, So this is why in my mind, OandM is not a separate thing. Its universal'' \end{quote}

Further needs described were being aware of their self-position while making turning moves (S13I, T5I) and   %interms of spatial awareness was how BLV people find it difficult to be aware of their self-position while making turning moves. 
%This was echoed in 2 focus groups by another 3 participants, both teachers and BLV students. 
%S13I shared, \begin{quote} ``When I was dancing, and you have to do these full circles. So one day, I didn't quite get myself twisted all the way around. I did side on to the audience, then somebody somehow managed to tell me%, because I'd done it wrong'' \end{quote}  
having better contrast in the environments where students with low vision learn and practice these lessons (S2I, S11I, S1F, S14F). %IS11 explained,\begin{quote} ``One of the really interesting things in that circus space that like dark rooms, dark floors, and the teachers, they've got short sleeves on. So the arms are usually quite white against the background. And so they're actually really well contrasted against the space. So that actually made a big difference to for actually being able to see them '' \end{quote} 
%A similar opinion was echoed in the focus groups as FS14 described, \begin{quote} ``So even establishments need to be just a little bit more understanding a little bit of contrast between the floor and the walls would be helpful for people you know, you can find the dark carpet over the light walls '' \end{quote}  
Safety was another spatial-related concern contributing to the lack of confidence to participate in physical activities from younger ages, as described by many teachers (T3I, T5I, T6I, T7I, T8I, T10I). %T8I shared,
%\begin{quote} ``They're very afraid of including students in physical education because they're worried about a student running into somebody, or someone running into the student for getting hit by a ball or tripping over something... 
%People focus on the safety, which is good. But people get so worried about safety that the person who could then sit out, 
%and that's actually really unsafe, for them to sit out, because it doesn't give them an opportunity to learn to use their body develop their balance, their strength.'' \end{quote}
%T6F and T9F echoed this concern in a focus group, and 
T6F questioned the researchers and other focus group participants whether technology can support in improving safety, \begin{quote} ``Would any technology to enhance safety be of benefit because I know quite a lot of vision impaired students are reluctant participants because of safety. An example might be fear of colliding with another person when playing cricket.'' \end{quote}

\subsubsection{Support system}
\label{sec:supportsystem}

BLV students find it useful to have a support person when participating in group sessions as shared by them and observed by teachers (S1I, S3I, S4I, S5I, S6I, S7I, S9I, S10I, S11I, S12I, T4I, T6I, T9I, T10I). Support people or sometimes peers and friends become guidance for BLV people in learning and accessing body movement by either describing what the teacher is performing or being the person next to them so that a low vision person can copy what they are doing, as S11I explains, \begin{quote} ``I look to the support workers to sort of mirror what they were doing, or sort of provide some guidance. As I was moving through the space, making sure that I was doing it, even not too badly '' \end{quote}

It was pointed out that the skill or experience of the support person can impact their learning. For instance, S3I mentioned that their support person was knowledgeable and helpful in learning a fast-paced environment. However, if the support person is not familiar with the particular body movement, it can result in learning inaccuracies, as S8I explained.

\subsubsection{Motivations}
\label{sec:motivations}

%(add quotes on needs of engaging with non-blind people, needed for the discussion points)

Among the motivations for learning body movement, the most frequently stated was social interaction (S1I, S3I, S4I, S6I, S7I, S9I, S10I, S12I). S10I explained their motivation to be included, \begin{quote} ``I'm just not very good at doing things at home. I don't have the motivation. I want to be a part of it, what I want is social interaction with other people.'' \end{quote} Some prefer attending a class with their companions who have inspired them, as IS1 mentioned when sharing about experimenting with different body movements, ``So I went with a friend... I started doing it online in lockdown just for fun with a friend who invited me...''. S3I stated how they intend to meet beyond BLV people by attending programs other than BLV-specific ones, ``%I have lots of friends who are blind or visually impaired, 
I don't just want to have contact with people who are blind and visually impaired. I don't want to be kind of so narrow that I don't have contact with other people''.

%BLV participants find the in-person learning more rewarding than online for social reasons. For instance IS4 mentioned \begin{quote} `` I've tried various online, you know, things, but I don't find it as rewarding, you know, because even though, you know, in a yoga studio, or dance studio or whatever, you're not necessarily socialising, you're still among other people. And that's part of the experience'' \end{quote} However, some situations have contributed to a lessening of motivation to attend body movement sessions as IS12 explained, \begin{quote} ``the bigger problems are the availability of sessions, the location of the sessions, and then building that pathway, to get the person from with vision impairment from their comfort zone into the session, and I don't mean to the front door, either I mean, into the session. And it wasn't that the people were unwelcoming, they didn't know how to accommodate a person who couldn't see, they didn't know how to make them feel welcomed, and make them feel part of the community '' \end{quote}

\subsubsection{Online Learning and Technologies}
\label{sec:online}
With the pandemic situation, rapid changes in learning environments were witnessed. This had an impact on body movement education and resulted in opportunities and challenges. 
S1I explained how her online teacher ``doesn't know how to set it up so she can see us'' performing while they are teaching. % \begin{quote} ``And with my friend who does Tai Chi, she can’t see any of us. She doesn't know how to set it up so she can see us. She needs to be watching herself '' \end{quote} 
Another BLV student (S13I) explained how she made use of selfie sticks and mobile technologies to ensure she is visible to her teacher.
The availability of accessible online content for a variety of sports was stated as a priority by S8I, \begin{quote} ``I think the most important thing is that there needs to be more online content for people with vision impairment with instructional content, on how to perform different sports. So, there need to be instructional videos on blind tennis, blind soccer.'' \end{quote} 

S2I (who has low vision and having further vision loss) pointed out that they had become used to accessing the same video as they don't find it as easy as other sighted people who can make use of their vision, \begin{quote} ``I guess the difference between me and someone who sighted is. I couldn't just scroll through Youtube and think, Oh, that one that's good. I'll do that one. I've got to kind of just do the ones that I'm used to '' \end{quote} 

%S2I further explained how they make use of videos by the support of their kids, \begin{quote} ``Then I might actually ask my kids to say my case. Hey, what's this lady doing? Can you show me what this idea is doing? And my kids are really good at this '' \end{quote} 

% IS1 shared their thoughts on having captions of learning videos, \begin{quote} `` Video plus audio but no audio description. But the thing is in the classes I am in, they are saying like an audio description. Now put your hands up, now breathe out. So its like the entire thing. I don’t even know what an audio describe would do, like she’s wearing pink, etc. She’s describing everything she’s doing. So its like purely listening to an audio description anyway '' \end{quote} 

%\subsubsection{Feedback}

Five BLV students (S4I, S5I, S6I, S7I, S9I) pointed out the lack of feedback as a key drawback of online learning including online classes and online content for BLV people. %S6I stated, ``you have a big, big problem is that you're not getting any feedback'' and 
S5I identified the lack of feedback as the reason for not using audio described online videos, ``[I have] No idea if I'm actually doing things correctly'' and stated they appreciate it if learning at home is made accessible, `` if you could get haptic feedback, and work on things in your own home with other described videos and stuff, that can be really useful''.

%S4I speculated that they would not imagine it to work, stating that: \begin{quote} ``It might tell me what they're doing, but it might not if I was trying to follow along that way. It probably wouldn't work '' \end{quote} 

%A similar speculation was made by IS9, \begin{quote} ``And you can you can say something, but how are they really realistically gonna show you? And I think my opinion is online classes will work the worst, because they cannot come and even touch you or the mean your over internet connection. They cannot see you, I guess you can try to put up yourself. But it's like, again, you have that class wide situation, right? She's up there, or he's up there. And okay, so how do I do this again? Oh, well, I'm too busy. I need to show the rest of the class. You know, there. There is 15 Other people beside you '' \end{quote}

%S7I also shared the concern of not being able to confirm what they are doing while online learning: \begin{quote} ``Now, you know, you can learn to do whatever online, yet, there is no substitute for that in-person learning that those little nuances of your hand is up and facing up instead of facing down'' \end{quote} 

%\subsubsection{Body movement elements}

\subsubsection{Body Awareness}
\label{sec:body}

Having the right body posture was said to be a key challenge, as prioritised in the focus groups. It was implied as a priority by nine participants in three focus groups. T10F shared their experience that they have to teach posture to many new BLV dancers, `` we find that quite a lot of the dancers, when they first come to us may not actually know how to walk properly...''.
%often when we're teaching to dance, we're also actually teaching them how to walk and posture how to stand so that they get the most economical use of their own body. '' \end{quote}
%A dance teacher (FT5) from another focus group shared a similar experience on posture: 
%\begin{quote} ``One of the things I'm always saying to students is, if you can't stand well, you can't move well, because it'll start to disintegrate. And one of the things I find with the blind students is a lot of them have sort of intrinsically bad posture.'' \end{quote}
Body awareness in terms of kinesthetics and proprioception (the ability of the body to sense the position, movement, and orientation of its own limbs and joints without relying on visual cues) was discussed by several teachers (T5I, T7I, T8I, T9I and T10I). T8F explained how posture is affected by kinesthetic awareness, \begin{quote} `` postural awareness and general body awareness, particularly for people who are congenitally blind, is very difficult because they don't have what all of us have, kinesthetic awareness. That kinesthetic awareness isn't supported by my eyes, looking at my hand, saying, Okay, this is where my hand is%, and that kinesthetic feel of where my hand is in space on the feedback you get from your muscles, my hand is up, or my hand is down 
'' \end{quote}

%T8 further explained in their interview how a student can reach mastery level by working on their kinesthetic awareness, ``when you start reaching that sort of that mastery level, you have a kinesthetic feel of how something should be the way the ball hits you... I just encourage them to tap into that as early as possible... that internal sense? What does that feel like, when you do it right?''

T5I who is experienced with teaching BLV adults, stated that BLV people have good proprioception skills. %\begin{quote} ``One thing I think that the blind students have got over the rest of us is they've got this very highly developed proprioceptive mechanism. So they know where every bit of their body is, at any given time, they don't have to be, you know, looking down at their feet, or looking in the mirror to see what's what's happening. And, you know, that's one thing they've got, you know, that's really good '' \end{quote}
However, T7I, who is more experienced with instructing BLV children talked about the lack of proprioception of congenitally blind students.
%\begin{quote} ``...first, you need to have a body consciousness or awareness. And that's, I know, when I'm throwing that my hand is after the behind me that is going besides my air, and then I stretch in a certain way, you see, often by blind students who are blind for a since birth, that they don't know exactly where their legs are, where their hands are, or what's, which speed they, they throw off'' \end{quote} 

%There is a quote by S11 or S14 on body lack of body coordination... it might be usefull to support the counter argument for just children... even T8 explains how some early development on o and m...

%\subsubsection{Body balance}

Another key challenge discussed along with posture was body balance. It was prioritised by four participants in three focus groups. Some dance teachers (T1I and T10I) avoid teaching movements such as turning around and jumping to BLV students as they impact body balance. %T10I stated,
%\begin{quote} ``we do find that it seems to be very common, their [BLV people] balance is not as good as a traditional dancer. So we can modify the routines so that they don't have to turn. %We definitely do not want a student to fall and we haven't had a student yet''\end{quote}
%S13F explained, "If you don't get the body posture right, that affects your balance " and 
%S12I stated, \begin{quote} ``Blind people are prone to balance problems, balance is very important. Blind people do have more falls than most people in the community'' \end{quote} From a teacher's (T8F) point of view, \begin{quote} ``It all comes back to balance for me. If you can't start from a balanced position, it's very difficult to teach anything more complex '' \end{quote}
S10 echoed their avoidance of any movement impacting balance in both interview and focus group discussion, ``I just ignore all the balance stuff, I can't do any of it''. %Particularly if I'm thinking about something like yoga, where they're saying, to focus on a spot in the distance, there's just no chance I can't do it''. %So just ignore it, or go and stand near a wall and hang on to a wall.
%'' \end{quote}
However, S2I had a different view on body balance where they believe body balance is about feeling rather than seeing.
%\begin{quote} ``You can feel, if you're balanced, or if you're not. This is something that you can be aware of without being able to see. %And actually, I think yoga is pretty good for that because they seem very in touch with how things should feel.'' \end{quote}
Based on a similar opinion of not relying on vision to balance body, T5F shared their experiences of how they apply their understanding of teaching BLV people to teach sighted students, encouraging them to focus on balance with closed eyes. %, \begin{quote} ``I've actually put into my classes with regular students, things like when they're balancing. Okay, close your eyes. Now see how we know when you haven't got that focus of the eyes to rely on and what other things in your body, have you got that you can use to help you balance. I sort of say to them,  my blind students balance, they can't see, they can't focus their eyes on a spot and try and stay there.'' \end{quote}
%FT8 explained factors affecting the balance of a BLV person interms of the onset of blindness and early childhood development,\begin{quote} ``One of the things we know when vision impairment happens to a person is really critical, whether they acquire that as an adult, or whether they're born with it as a child, people who have been born with vision impairment, or who have acquired vision impairment very early in life, the development that they go through in terms of, you know, learning to crawl and learning to walk and those sorts of things. It happens along the same pathway. But because there's less stimuli visually, because there's a lot more challenge to balance because you don't have the visual input...'' \end{quote}
T2F expressed that a cane might be provided if the person needs a third point of contact on the ground to balance themselves. However, if the balance is caused by reasons that are not vision-related they would recommend meeting a physiotherapist.
%\begin{quote} ``But if I have to make a judgement about whether this is just a vision related, or whether there are other neurological problems, ageing problems, that might be causing that imbalance, because if that's the case, straight to the physiotherapist, who's more qualified to do that deeper investigation of what might be causing the balance issues'' \end{quote}
In the same focus group, T9F expressed how body balance is challenging when other elements of body movement come into play, such as weight transfer and speed.
%\begin{quote} ``Balance alone is not the problem, it becomes a problem when weight transfer and fast pace comes in. %I think balance both for myself and for a lot of people I work with it becomes an issue when it becomes part of weight transfer, when you're doing a fast movement, and moving from one to the other and that's when you can be a bit off balance is probably the only way. So I don't say balance directly. It's more of a part of other elements that I would say it as.
%'' \end{quote}

\subsubsection{Movement Qualities}
\label{sec:qualities}

Some BLV students expressed that understanding weight transfer (S6F and S13F) is a challenge and that it is intertwined with balance. %S13F stated, "I find the weight transfer thing quite difficult. When I'm trying to do weights and my instructor will say to me, you need to bend your knee a little bit more, or you need to balance more on both legs. It takes a while to sort of get to get yourself completely balanced and then having enough tension in your arms actually to pull up a weight". The learning of where the force should originate to expect a certain body movement target is said to be challenging as described by FT9:
% \begin{quote} ``thing that we have trouble with is a lot of people think that it comes from your arms and your shoulders, and it actually doesn't. Force actually comes from your hip and your legs. I guess when it comes to force, a couple of things that we have trouble sort of teaching and explaining of how to get those movements actually to get the force in the end.'' \end{quote}
T9F explained how they find it challenging to teach blind students how much force or how fluid the movement should be. 
%\begin{quote} ``%Being someone who's partially sighted, you know, I might need to stand a bit closer, but I still can watch someone and see what movement looks like. 
%When I'm teaching I actually find people with no vision, they actually tend to be quite stiff and tighten up the muscles rather than moving being quite fluid is what I notice '' \end{quote}
%\subsubsection{Timing and Speed}
A key challenge was discussed and prioritised in one focus group (by three teachers) on how to teach fluidity of movement by combining speed and application of force. T6F raised the problem by explaining, \begin{quote} ``for example, in an efficient running stride, you might have students mapping you as in feeling your upper and lower leg as you talk about the movement, but they're feeling it in slow motion, they're not feeling it in actual time... %demonstrating a faster movement such as a running stride in slow motion, there's a few gaps, and undisciplined, 
wondering how to bridge those gaps and make it more fluid. '' \end{quote}

%Agreeing to the same challenge, T9F added their experiences to it, "we'll use body mapping a lot, particularly if I'm struggling to get my words out and actually describe it sometimes not always. But it does tend to slow down and the movement isn't fast, I would just agree if there is a way to speed it up and make it more fluid and what it's meant to be, rather than having to do a demonstration of it".

In the same discussion, another teacher (T2F) suggested an idea to address this problem of fluidity teaching, ``I wonder if it's possible to develop something that gives like a positive audio cue when the body's getting it right''. %Because really, when vision gives you very instant feedback if you're moving very awkwardly, it looks untidy and ungainly 
%'' \end{quote}

In the interviews, BLV people expressed that they were comfortable understanding the timing of movement, and it was not discussed as a problem. However, it was raised and stated (without prompting) otherwise in two focus groups by two BLV students and three teachers. S10F shared, %\begin{quote} ``%for me, 
something that I think I struggle with is timing. So if it's something that has a movement, that is repetitive, like steps in a dance, like, the actual timing of the moves is often lacking''. %\end{quote}
S7I shared an idea of how wearables can be integrated to learn body movement, including timing and speed of it, ``if you're trying to do fast punches with taekwondo or something, maybe having it set up so... %that tap for, right, you have like two wristbands on and like, tap, tap. So it's telling you which hand should be punching, and 
it could be speeding up to do it [tapping] quicker or slowing down''. %\end{quote}

\subsubsection{Summary}
The ten themes described above provide insights into the key challenges and needs of BLV students and teachers. There are several expectations of BLV students on verbal instructions. However, teachers find it challenging to rely solely on verbal instructions and require other techniques and tools supporting in addition to verbal instructions. While physical guidance plays a supporting role in educating BLV students, there are some indications of hindrances to physical contact. Key problematic areas of teaching BLV students were around body awareness, spatial awareness and movement qualities. BLV students indicated the value of online learning. However, they shared some concerns and needs on accessing online content. Community sharing of expertise was suggested in the interviews and was observed in practice through focus groups. The BLV students expressed the need for human interactions in terms of support and social engagement.

\section{Discussion}
\begin{figure*}[ht]
  \centering
  \includegraphics[scale=0.28]{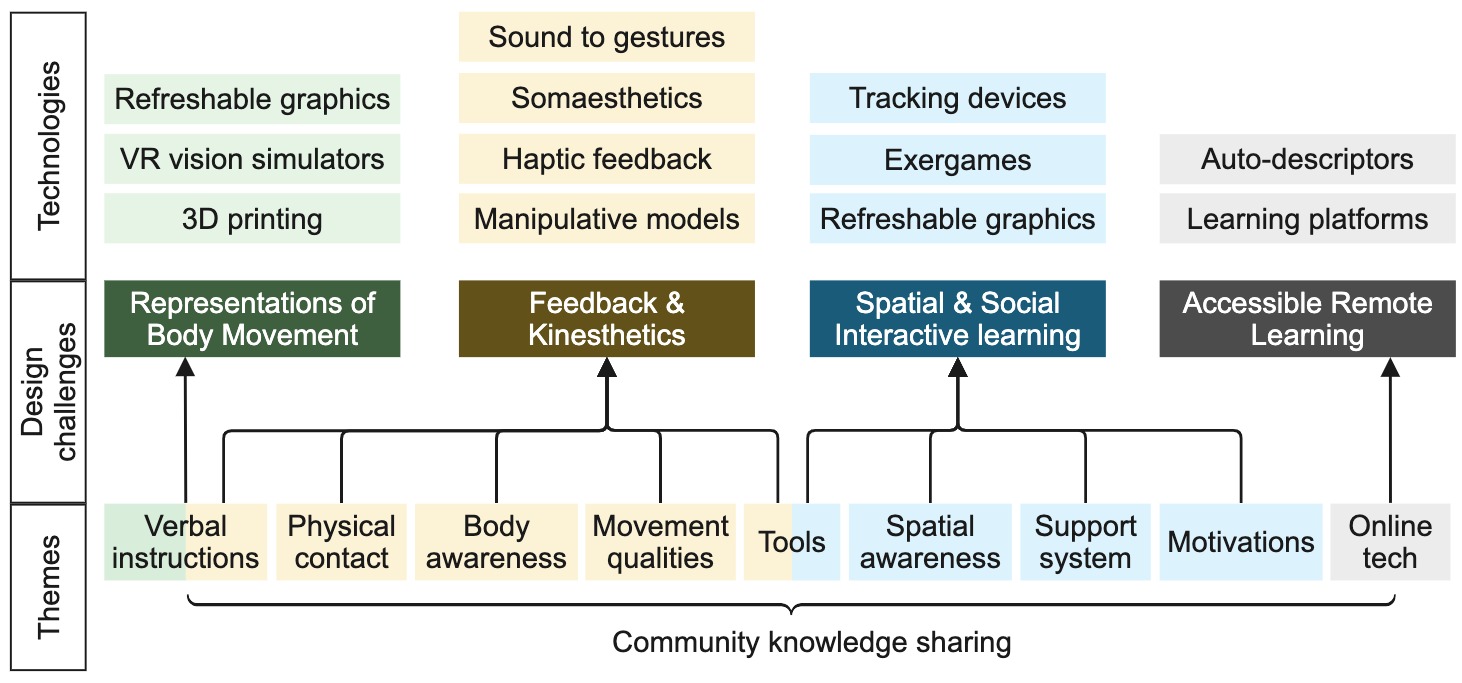}
  \caption{Mapping of themes from results to key design challenges and technology implications.}
  \Description{This is a mapping of themes from results to key design challenges and technology implications. There are three levels, from the bottom to the top. Themes are listed at the bottom linking to the relevant derived design challenge on top of them. Then the four design challenges are linked to potential technologies. Verbal instructions theme leads to, the first design challenge of the representation of body movement to support verbal instructions. This then leads to technologies of 3D printing, VR vision simulators and refreshable graphics. Themes on verbal instructions, physical contact, body awareness, movement qualities and tools lead to, the second design challenge of supporting feedback and improving kinesthetic awareness. This then leads to technologies of manipulative models, haptic feedback, somaesthetics and sound to gestures. Themes on tools, spatial awareness, support systems and motivations lead to, the third design challenge of supporting spatial and social interactive body movement learning.  This then leads to technologies of refreshable graphics, exergames and tracking devices. The theme of online learning and technologies leads to, the fourth design challenge of supporting accessible remote learning. This then leads to technologies of learning platforms and automatic descriptors. The theme of community sharing leads to all design challenges and thus to all technology implications.}
  \label{fig:Themes}
\end{figure*}

In this section, we discuss the summary of the insights gained from all three studies and identify four key design challenges (Figure~\ref{fig:Themes}). Each key challenge is defined based on our findings and is linked to past literature providing implications for technological opportunities.

\subsection{Design Challenge 1: Representation of Body Movement to Support Verbal Instructions}

 One of the major concerns repeatedly voiced by BLV people is the lack of detailed body movement instructions. However, describing a body movement purely in verbal terms is a complex task for teachers. This indicates design challenge 1: exploring supplementary mediums for verbal instructions supporting teachers to convey body movement details that could also act as a second mode of confirmation for BLV students.
 
A compounding problem, pointed out by a teacher with low vision, is that there are diverse experiences of blindness. Some teachers try to empathise by using different techniques such as closed eyes instructions (section~\ref{sec:verbalinstructions}). There is potential to explore the solution space on evolving use of vision simulation ~\cite{tigwell2021, goodman2007} based on virtual reality ~\cite{yao2022} and wearable displays ~\cite{zhang2022, ates2015}. However, it is important to be mindful of problems with relating to a BLV person without lived experiences ~\cite{bennett2019,french1992}. 
 
Some teachers and studies~\cite{rector2017} suggest the use of metaphors when describing movement, which was a strength for low vision students and blind students who lost their vision in later stages of life. However, congenitally blind students find it difficult to relate to the visual aspects of such metaphors (section~\ref{sec:verbalinstructions}). Furthermore, it was pointed out that there are vocabulary issues when teachers refer to the same pose by different names. Several teachers and some BLV students shared that BLV people require a second mode of confirmation, which sighted students get through visual stimuli. As some BLV students suggested, a collection of body movement representations similar to a dictionary in tactile format~\cite{Phutane2022} such as 3D printed moving graphics~\cite{kim_toward_2015}, static or dynamic tactile graphics on refreshable displays~\cite{guinness2019, holloway2022} could be potential solutions to explore. For instance, Kim and Yeh~\cite{kim_toward_2015} created 3D printed movable tactile pictures for BLV children where they suggest using a movable set of components that can be added to 3D prints to represent the motion and understand the kinetic experiences, which can be further explored to represent human body movement. 

Another instance of using tactile graphics is the exploration conducted by Holloway and colleagues~\cite{holloway2022} that indicates the potential of using refreshable graphics displays to represent body movement concepts. 
%However, they indicate a limitation of showing multiple limb movements. 
Similarly, Kobayashi and Tatsumi~\cite{kobayashi2021} have explored refreshable tactile graphics to represent Floor-Volleyball motion to BLV players, and their results indicate a complication of showing the shape of players. 
%However, they indicate the potential for representing simplified motion and speed of the dominant wrist. 
While both works highlight some limitations in representing whole body movement, refreshable tactile displays demonstrate other strategies for representing shapes and poses in tactile graphics. %Although much research has aimed at creating tactile materials for STEM subjects, less attention has been given to their effectiveness in classroom environments~\cite{Phutane2022}. 
Furthermore, some research has been conducted on using tactile materials to support orientation and mobility concepts~\cite{Palivcova2020} but little research considering details of body movement and contextualised physical activities such as dance and sports ~\cite{Butler2021, kobayashi2021}. %Here we reiterate the gap of research in technologies conducted to support the communication between the teacher and the student, as the majority of studies have been conducted either to facilitate participating in an activity without a sighted guide or to improve performance independently. 

\subsection{Design Challenge 2: Supporting Feedback and Improving Kinesthetic Awareness}
A majority of teachers make use of physical contact or tactile instructions in addition to verbal instructions when adjusting posture and providing feedback to BLV people~\cite{oConnell2006}. However, there are several social factors that can inhibit the use of physical contact (section~\ref{sec:physicalcontact}). Furthermore, a major challenge highlighted by teachers is that they find it difficult to explain the fluidity of movement, even through physical manipulations (section~\ref{sec:qualities}). By fluidity, they mean kinesthetic awareness of how fast the movement should be, how much force to apply and from which part of the body it should be applied or supposed to be feeling. Thus, this imposes two design challenges. One is to explore how we can imitate the benefits of physical contact. The second is to explore how the fluidity of body movement can be communicated to BLV people. 

Some teachers specialising in training children and younger adults use manipulative human models such as dolls and figures to provide tactile feedback (section~\ref{sec:tools}). Literature indicates the wide use of 3D models for educational purposes~\cite{ford2019}, medical science contexts for understanding human anatomy~\cite{nadagouda2020, ye2020}, and some early experiments of using humanoids~\cite{lapeyre2014,saini2022} with little to no research on the use of 3D human models in the physical education contexts for BLV people. %~\cite{simpson2021}. 
Further, some teachers use and recommend sound-based tools to convey the fluidity of movement and provide feedback. The preliminary survey reported much less use of sound-based tools by teachers (section~\ref{sec:surveylearningteachingmethods}), with the tools being mostly low or not technology-oriented (whistles, beeping balls, clapping). However, as pointed out in interviews (section~\ref{sec:tools}), using the recorded sounds of movement to provide feedback to BLV students is a strength. None of the adult BLV participants reported being exposed to the use of 3D model figures and sound-based tools. These findings motivate us to investigate how manipulative human models~\cite{claire2023} and sound-based tools~\cite{bergsland2015, dias2019, ilsar2020, mandanici2020, ramsay2020, sadasue2021} can be used as a technique by the wider teaching community to provide feedback and represent body fluidity for BLV learners. For instance, our findings highlight the importance of considering fluidity of movement, especially in  goalball throws, beyond understanding the direction of a moving ball~\cite{miura2018, watanabe2022}. 

 %This raises the question of exploring further how such equipment could be adapted in adult classes or in a body movement class setting showing characteristics of fast-paced and several students to be attended. Further, survey respondents pointed out the practical concerns of creating 3D models and using sound-based tools (sensory impact on others). Moreover, teachers' use of such models was discussed mostly in an individual setting rather than teaching as a group.

An insight shared by BLV participants is that they valued knowing how and where they are supposed to feel when engaging in a certain body movement (section~\ref{sec:verbalinstructions}). Similarly, teachers shared the importance of kinesthetic awareness impacting body posture and balance (section~\ref{sec:body}). This inspires us to explore how concepts such as somatic pedagogy~\cite{seham2015} and somaesthetics can facilitate in conveying the sensory experiences of body movement~\cite{hook2021}. For instance, Seham and Yeo \cite{seham2015} state, ``Although these methods have not been empirically studied nor specifically designed for people with visual impairment, somatic pedagogy presents viable, non-visual movement communication strategies compatible with the learning needs of blind students''. Further, imitating physical contact or touch simulation to improve kinesthetic awareness can be explored in terms of haptics-based ideas~\cite{song2015}. Haptic feedback-based solutions have been explored to understand whole-body navigation and particular limb movement~\cite{hong2017}. However, there is less research conducted on fine-tuning the quality of movement with the intention of teaching body movement and supporting the connection between the teacher and BLV learner~\cite{aggravi2016, malik2021}. For instance,  Malik and colleagues~\cite{malik2021} presented a novel technology design by developing an algorithm for a sensor-based mat to support blind people to be in sync in a group-based aerobics class. While this supports tracking the time of the steps, which is a key aspect of quality of movement, there is an underlying assumption that the learner is familiar with the exercises. They have also explored reactive verbal feedback, which is a key need our studies identify as well. It can be beneficial to explore the potential of other kinds of feedback methods, consider the movement of different parts of the body, and focus on other spatial arrangements beyond a confined area of mat space.

%Some BLV participants who are interested in dance explained the difficulties of understanding dance conceptually and learning practically. Some dance teachers make use of tactile or physical contact to convey patterns and steps in a movement. Further, very little recent research has been done on teaching dance steps with technology supporting BLV people [Dias2019]. However, the teachers expressed difficulty of explaining qualities and details of creative dance movements, which are not aimed by past research work. Some studies have aimed at representing body movement such as sport events for BLV audiences as entertainment [Ohshima 2021, ~\cite{goncu_did_2021} and very few on non-visual representations of creative movement such as dance [Bläsing 2021, Ravenscroft 2019]. However, there is a design challenge to be explored on how creative body movement could be represented for BLV people beyond as an audience and in a learning perspective. 

% \subsection{Design challenge 3: supporting body balance}

% XXX I just thought this now... because there was alot of discussion around body balance, coordination and how it is impacted by other things or how that is the basis for everyother thing (results on body movement elements...) because this is something relatable to falls, and thereby safety.. I saw some papers on body balance but not particularly while engaging in other physical activities.. need to check more on current literature...

\subsection{Design Challenge 3: Supporting Spatial and Social Interactive Body Movement Learning}

 Orientation and Mobility training improve spatial awareness and navigation of BLV people. Additionally, support workers and other companions provide assistance to BLV people. 
 One major concern raised on spatial understanding was reduced awareness of others in the space and their movement (section~\ref{sec:spatialsocial}), which we present as a design challenge to be addressed. This was also presented as a reason contributing to safety and confidence. Another insight from the findings was that social interaction is a key motivation (section~\ref{sec:motivations}) for BLV people to participate in body movement activities and attend in-person classes. Interestingly, BLV people showed interest in social engagement and were less concerned about independence when participating in body movement. This insight shows the potential for another design challenge to explore the combination of the needs of BLV people for social awareness and social interaction.

There are some attempts by teachers to represent and train to respond to game changes, such as player positions using tools such as magnetic boards (section~\ref{sec:tools}). Refreshable tactile graphics~\cite{kobayashi2021, holloway2022} could also be potential solutions. However, those require the game to be paused in the middle or be moved away from the court, interfering with the spontaneity of the activity. Research has widely explored supporting BLV people in navigating outdoor~\cite{swaminathan2021, huber2022, anthierens2018}, or indoor spaces~\cite{amemiya_orienting_2010, jeon_listen2droom_2012, ko_situation-based_2011, Palivcova2020}, and object tracking~\cite{he_pneufetch2020, richardson2022, cooper2022}. Furthermore, there are studies supporting directional guidance as advised by instructors~\cite{aggravi2016}. However, the needs raised from this study are focused towards the awareness of the movement of other people in a space and in a fast-moving pace. 

Many studies involving different technologies have aimed at real-time person tracking~\cite{wu2017, tan2019, gai2019} including collocated person tracking in virtual reality ~\cite{weissker2021} supporting collision avoidance ~\cite{scavarelli2017}. Some studies have explored person tracking aiming for BLV people in applications such as tracking people standing in lines ~\cite{Kuribayashi2021}, and tracking for distance maintenance in a crowd due to Covid-19 ~\cite{shrestha2020}. However, there is less evidence of the exploration in the context of body movement learning spaces. Al-Zayer, Folmer and colleagues~\cite{al2016, folmer2015} conducted a Wizard of Oz study investigating how the sound cues of drones can support blind runners to stay in line in an indoor space without obstacles and the need for a sighted guide. While safety is a key spatial concern that our studies identified, attention is also needed on ways to be aware of themselves in a space shared with many others. Furthermore, support from another person was indicated as a strength, especially in improving human interactions. Solutions addressing this need could result in supporting BLV people to engage in group-based body movements such as group sports and group dance performances, opening more avenues for inclusion. %One other need this is addressing is how BLV people can co-exist in an in-person classroom full of people irrespective of blind-specific or not (refer to results section). Exploring solutions to these needs could also indirectly support the previous challenge on social interactive body movement learning.
%Social interaction is the most popular reason for body movement participation. It is one of the key motivations for many BLV students to attend in-person body movement classes. Some pointed out how demotivated they are to learn on their own at home or online. This creates the design challenge to explore how those social motivations can be further fulfilled with the support of technology-based means to encourage participation in offline or online body movement learning programs. %As pointed out by T8 (refer to section), solutions can be explored that can facilitate the same skills that are missed due to less inclusive/adaptable sports or other activities. 
Further adding to the second design challenge, research has been explored on making physical play such as exergames accessible to address the needs of BLV people ~\cite{morellivi-bowling2010, morellivi-tennis2010, morelli_hapticaudio_2010, gasperetti2010}. However, a few such tools let BLV participants interact socially with others online, while the focus on in-person social interactions is minimal ~\cite{battistin2023}.  

%\subsection{Design challenge 5: supporting navigation in confined spaces with other people}

\subsection{Design Challenge 4: Supporting Accessible Remote Learning}

With the Covid-19 situation, people relied on remote learning programs to practice or learn body movement, which could also benefit BLV people to overcome various other barriers, including travelling and finding accessible in-person classes~\cite{Lee2014, jaarsma2014}. However, remote learning is said to be lacking feedback even more than in-person activities (section~\ref{sec:online}). Feedback-focused research studies conducted for accessible remote learning of physical activities are limited ~\cite{rector2017, dias2019}. For instance, Dias and colleagues~\cite{dias2019} suggest an asynchronous platform that provides auditory feedback when a student attempts a pre-recorded dance lesson, and Rector and colleagues~\cite{rector2017} have designed a solution providing auditory confirmation for blind yoga learners at home. The research community could explore accessible synchronous learning strategies and experiment with other modalities, including haptics, as suggested by our participants (section~\ref{sec:online}). Exergames such as VI-bowling~\cite{morellivi-bowling2010} and VI-tennis~\cite{morellivi-tennis2010} have adapted sports games by converting visual cues to audio and haptic feedback. Although improving energy expenditure has been the primary aim of these studies, tools and techniques introduced in these works have further potential to be explored for motor learning. For instance, VI-bowling~\cite{morellivi-bowling2010} uses a technique that can point out a direction to a player, tactile dowsing, which could be explored for accessible learning of physical activities. 

Furthermore, there is a need to focus on supporting the creation of accessible online content, especially videos with descriptions (section~\ref{sec:online}). %Some BLV people referred to online materials even to be included in in-person group spaces by preparing ahead and becoming familiar with body poses. 
Several studies have focused on video accessibility, such as providing automatic descriptions ~\cite{wang2021, filho2021}, including humans in the loop with machine learning descriptions ~\cite{yuksel2020}, collaborating with sighted writers ~\cite{natalie2021} or BLV writers ~\cite{jiang2022}, querying supported bots `InfoBot's ~\cite{bodi2021}. These studies indicate the descriptions of characters and actions. However, for the objective of learning body movement, further studies are essential to inquire about the needs of BLV people when deriving descriptions to create accessible online content. Additionally, as pointed out by our participants, searching options should also be considered when considering access to online content. %For instance, some describing the character's clothing or the scene might not be the most important while the attention should be more on the body action.   
%\subsection{Design challenge 7: Supporting the understanding of creative body movement}

\section{Limitations and Future Work}
In our work, we intended to understand the challenges of people with lived experiences in terms of our participants. Thus, we aimed for teachers that have experience in teaching BLV students. A further contribution is possible if the accessibility research community can explore and design solutions considering both expert and novice teachers or teachers with no prior experience in educating BLV people. Our findings do not represent the needs of all BLV people, and we encourage researchers to explore further barriers to body movement when designing technological solutions. Additionally, there could be an impact on the results based on the varied experience of teachers, including their length of experience, the number of BLV students they taught and the blindness variations of their students. Furthermore, we intended the focus group participants to prioritise key challenges by selecting and ranking them. Although participants suggested some as key challenges, they found it difficult to choose one above the other and implied that these challenges are dependent on the variation of blindness. However, having both individual perspectives through interviews and group perspectives through interactive focus group discussions contributed to the identification of validated insights. 

\section{Conclusion}

In this paper, we describe the findings of a preliminary survey study, semi-structured interviews involving thirteen BLV students and ten teachers, and four focus group studies  involving seven BLV students and six teachers to understand their current body movement educational techniques and related challenges. Our work identifies ten insightful themes and four key design challenges to be addressed, proposing potential areas of solutions extending assistive technologies research on accessible representations, kinesthetic awareness, spatial awareness, and remote learning support for body movement education of BLV people. %Further, we found that social engagement and support play a bigger role in body movement participation, with less concern for independence. 
Finally, we encourage the assistive technologies community to co-design potential solutions to these identified challenges promoting the quality of life of BLV people and supporting the teachers in the provision of inclusive education.

%\section{Acknowledgments}

%%
%% The next two lines define the bibliography style to be used, and
%% the bibliography file.
\bibliographystyle{ACM-Reference-Format}
\bibliography{sample-base}

%%
%% If your work has an appendix, this is the place to put it.
%\newpage
\appendix

%\section{Appendices}
\clearpage
\onecolumn
\section{Individual Demographic Information of Interview and Focus Group Participants}

\begin{table*}[ht]
\begin{tabular}{lllllllll}
\hline
\textbf{ID} & \textbf{Age} & \textbf{Country} & \textbf{G} & \textbf{B} & \textbf{Onset} & \textbf{Tech Adoption} & \textbf{Body Movement} & \textbf{F} \\ \hline
S1 & 45-54 & Australia & F & L & Acquired before 20yrs & Need support & D,S,M,Y & Y \\
S2 & 35-44 & USA & F & L & Congenitally blind & Up-to-date & D,Y & - \\ 
S3 & 55-64 & Australia & F & L & Congenitally blind & Up-to-date & S,M,Y & Y \\ 
S4 & 65-74 & USA & F & L & Acquired before 20yrs & Up-to-date & D,M,Y & - \\ 
S5 & 35-44 & Australia & F & T & Congenitally blind & Early adopter & D,S & - \\ 
S6 & 55-64 & Australia & M & T & Congenitally blind & Up-to-date & S,M & Y \\ 
S7 & 55-64 & USA & F & T & Acquired before 20yrs & Need support & D,S,M & - \\ 
S8 & 35-44 & Australia & M & L & Acquired before 10yrs & Up-to-date & S,M & - \\ 
S9 & 25-34 & USA & F & T & Acquired before 10yrs & Early adopter & S,M & - \\ 
S10 & 45-54 & Australia & M & T & Acquired before 40yrs & Early adopter & D,S,M & Y \\ 
S11 & 35-44 & Australia & N & L & Congenitally blind & Early adopter & S,C & Y \\
S12 & 75-up & Australia & F & T & Acquired before 60yrs & Need support & D,Y & - \\ 
S13 & 55-64 & Australia & F & T & Congenitally blind & Early adopter & D,Y & Y \\ 
S14* & 45-54 & Australia & F & L & Acquired before 30yrs & Early adopter & D,S,Y & Y \\ \hline
\end{tabular}
\caption{Demographics and other information of BLV students participated in the interviews. `Y' stands for participants who joined the focus group discussions (F). *S14 participated in focus group discussions only. Different body movement categories are identified as `D' for dance, `S' for sports, `M' for martial arts, `Y' for yoga, and `C' for circus. Blindness (B) categories are referred by `T' is for totally blind, and `L' is for legally blind or low vision. Gender (G) is listed by referring to female as ‘F’, male as ‘M’ and non-binary as ‘N’.}
  \label{tab:blvParticipants}
\end{table*}

\begin{table*}[ht]
\begin{tabular}{llllllllll}
\hline
\textbf{ID} & \textbf{Age} & \textbf{G} & \textbf{\begin{tabular}[c]{@{}l@{}}Expertise \\ (type)\end{tabular}} & \textbf{\begin{tabular}[c]{@{}l@{}}Expertise \\ (style)\end{tabular}} & \textbf{\begin{tabular}[c]{@{}l@{}}Expertise \\ (yrs)\end{tabular}} & \textbf{\begin{tabular}[c]{@{}l@{}}Taught \\ (age)\end{tabular}} & \textbf{\begin{tabular}[c]{@{}l@{}}Taught \\ (\#)\end{tabular}} & \textbf{Tech Adoption} & \textbf{F} \\ \hline
T1 & 35-44 & F & D & West African dance & 22 & A & 50 & Up-to-date & - \\
T2 & 55-64 & F & O\&M & - & 29 & C,A & 200-300 & Need support & Y \\
T3 & 45-54 & M & D & Tango & 5 & C,A & Many & Up-to-date & - \\
T4 & 25-34 & F & D & Commercial Jazz & 12 & C,A & 3 & Up-to-date & - \\
T5 & 65-75 & F & D & Ballet & 10 & A & Around 12 & Up-to-date & Y \\
T6 & 45-54 & M & D,S,PE & Many & 25 & C,A & 80 & Up-to-date & Y \\
T7 & 55-64 & M & S,PE & Many & 36 & C & Around 100 & Need support & - \\
T8 & 35-44 & N & S,PE & Many & 18 & C,A & 200 & Early adopter & Y \\
T9* & 25-34 & F & S & Goalball & 10 & C,A & Unsure & Up-to-date & Y \\
T10 & 55-64 & M & D & Ballroom & 2 & A & 16 & Need support & Y \\ \hline
\end{tabular}
\caption{Demographics and other information of teachers who participated in the interviews. `Y' stands for participants who joined the focus group discussions (F). Different body movement expertise categories are identified as `D' for dance, `S' for sports, `PE' for physical education, and `O\&M' for Orientation and Mobility. Gender (G) is listed by referring to female as ‘F’, male as ‘M’ and non-binary as ‘N’. `C' indicates teachers who have taught children and `A' who have taught adults. *All teachers are sighted except T9 is legally blind and acquired at before 30 years. All teachers are from Australia except for T3 from Egypt and T7 from the Netherlands.}
  \label{tab:teachers}
\end{table*}

\clearpage
\onecolumn
\section{Themes of interview and focus group discussions analysis}

\begin{table*}[ht]
\begin{tabular}{@{}llllllll@{}}
\toprule
\textbf{Theme} & \textbf{Sub theme / code} & \textbf{I} & \textbf{S} & \textbf{T} & \textbf{F} & \textbf{S} & \textbf{T} \\ \midrule
Verbal instructions & Difficulties in explaining details & Y & Y & Y & Y & Y & Y \\
Verbal instructions & Relatable examples and counter argument & Y & Y & Y & Y & Y & - \\
Verbal instructions & Empathised teaching & Y & Y & Y & Y & - & - \\
Verbal instructions & Vocabulary needs & Y & Y & - & Y & Y & - \\
Verbal instructions & Concentration & Y & Y & - & - & - & - \\
Verbal instructions & Need of additional modalities & Y & Y & Y & Y & Y & Y \\
Verbal instructions & Feel of movement & Y & Y & - & - & - & - \\
Physical contact and guidance & Benefits of physical contact based methods & Y & Y & Y & Y & Y & Y \\
Physical contact and guidance & Consent and preferences for physical contact & Y & Y & Y & Y & Y & - \\
Physical contact and guidance & Impact of the Covid-19 & Y & Y & Y & - & - & - \\
Physical contact and guidance & Inherent physical contact & Y & Y & Y & Y & Y & Y \\
Tools supporting teaching & 3D modals, tactile diagrams, recording sound & Y & Y & Y & Y & Y & Y \\
Community sharing and practice & Professional bodies and consultations & Y & - & Y & Y & Y & Y \\
Community sharing and practice & Different experts' perspectives & - & - & - & Y & - & Y \\
Spatial and social awareness & Awareness of others & Y & Y & Y & Y & Y & Y \\
Spatial and social awareness & Turning moves & Y & Y & Y & Y & Y & Y \\
Spatial and social awareness & Needs on contrast & Y & Y & - & Y & Y & - \\
Spatial and social awareness & Safety concerns & Y & Y & Y & Y & - & Y \\
Support system & Benefits of support persons & Y & Y & Y & - & - & - \\
Support system & Copying another & Y & Y & - & - & - & - \\
Motivations & Social reasons & Y & Y & - & - & - & - \\
Motivations & Impact of Covid-19 & Y & Y & - & - & - & - \\
Online learning and teaching & Expertise with using new technologies & Y & Y & Y & - & - & - \\
Online learning and teaching & Online content & Y & Y & - & - & - & - \\
Online learning and teaching & Lack of feedback & Y & Y & - & - & - & - \\
Body awareness & Body posture & Y & Y & Y & Y & Y & Y \\
Body awareness & Kinesthetics and proprioception & Y & Y & Y & Y & - & Y \\
Body awareness & Body balance & Y & Y & Y & Y & Y & Y \\
Movement qualities & Force and weight transfer & Y & Y & Y & Y & Y & Y \\
Movement qualities & Speed and timing & - & - & - & Y & Y & Y \\ \bottomrule
\end{tabular}
\caption{Themes emerged from interview study and focus group study analysis. `I' represents interviews, `F' represents focus groups, `T' represents teachers, `S' represents BLV students and `Y' indicates in which study the theme emerged.}
  \label{tab:themes}
\end{table*}

\end{document}